\documentclass[12pt]{article}
\usepackage{amsmath, amssymb, amsfonts}
\usepackage{siunitx} 
\usepackage{graphicx}
\graphicspath{{graphs/}}
\usepackage{booktabs}
\usepackage{longtable}
\usepackage{geometry}
\usepackage{subcaption}
\usepackage{enumitem}
\usepackage{xcolor}
\usepackage{float}
\usepackage{url}
\usepackage{threeparttable}
\usepackage{natbib}
\usepackage{pdflscape}
\usepackage{float}
\usepackage{subcaption}
\usepackage{graphicx}
\usepackage{hyperref}
\usepackage[normalem]{ulem}
\newcolumntype{A}{S[table-format=2.4]} 
\newcolumntype{B}{S[table-format=4.0]} 
\newcolumntype{C}{S[table-format=1.5]} 
\newcolumntype{D}{S[table-format=3.2]} 
\newcolumntype{E}{S[table-format=5.2]} 

\newtheorem{definition}{Definition}

\numberwithin{figure}{section}
\numberwithin{table}{section}

\newcommand \revision[1]{\textcolor{black}{#1}}

\title{Trading Electrons: Predicting DART Spread Spikes in ISO Electricity Markets}
\author{Emma Hubert\thanks{CEREMADE, Université Paris Dauphine \& PSL Research University.}\and Dimitrios Lolas\thanks{Department of Operations Research \& Financial Engineering, Princeton University.} \and Ronnie Sircar\footnotemark[2]}
\date{Revised \today}

\begin{document}
\maketitle
\begin{abstract}
We study the problem of forecasting and optimally trading day-ahead versus real-time (DART) price spreads in U.S. wholesale electricity markets. Building on the framework of \cite{galarneau2022foreseeing}, we extend spike prediction from a single zone to a multi-zone setting and treat both positive and negative DART spikes within a unified statistical model. To translate directional signals into economically meaningful positions, we develop a structural and market-consistent price impact model based on day-ahead bid stacks. This yields closed-form expressions for the optimal vector of zonal INC/DEC quantities, capturing asymmetric buy/sell impacts and cross-zone congestion effects. When applied to NYISO, the resulting impact-aware strategy significantly improves the risk–return profile relative to unit-size trading and highlights substantial heterogeneity across markets and seasons.
 
\end{abstract}

\section{Introduction} 

In U.S. wholesale electricity markets operated by Independent System Operators and Regional Transmission Organizations (ISOs/RTOs), trading is organized as a two-settlement system: a Day-Ahead Market (DAM), in which schedules and prices for the following operating day are determined, and a Real-Time Market (RTM), in which actual imbalances are settled at higher frequency. The difference between the two prices, the day-ahead real-time spread (DART), is a central risk factor for both financial and physical market participants. 

\revision{Since the Day-Ahead Market clears on the basis of forecasts of supply, demand, and network conditions, day-ahead prices can be thought of as expectations of real-time prices formed the day before delivery. The inherent uncertainty in these forecasts, stemming from demand variability, generator availability, and transmission conditions, implies that day-ahead and real-time prices will systematically differ. These deviations can be particularly large in organized U.S.\ wholesale electricity markets, where ISOs and RTOs clear day-ahead and real-time markets through security-constrained economic dispatch under the regulatory framework overseen by FERC: since locational marginal prices reflect the marginal cost of serving load at each location, including the effects of congestion and other operational constraints, forecast errors in demand or network conditions can translate into substantial price deviations between the two settlements. Physical market participants such as generators and load-serving entities are exposed to the risk of such deviations and may be willing to pay a premium to hedge against them, embedding a risk premium in day-ahead prices. The DART spread therefore reflects both a market-implied forecast error and this embedded risk premium \cite{longstaff2004electricity}, which can persist even in an efficient market as compensation for bearing short-term pricing risk. Large deviations can thus generate substantial profit opportunities for traders capable of anticipating extreme DART events.}

In NYISO (New York ISO), ISO--NE (ISO New England) and ERCOT (Electric Reliability Council of Texas), market participants may take purely financial day-ahead positions through virtual bidding \cite{hogan2016virtualbidding}. \revision{In these markets, virtual trading relies on two instruments: incremental (INC) and decremental (DEC) bids, named after their effect on the aggregate supply and demand curves submitted to the Day-Ahead Market.}
{\color{black} \begin{definition} 
Let $P^{\rm RT}$ denote the real-time price, and $P^{\rm DA}$ the corresponding day-ahead price, both in \$/MWh. The {\rm DART} spread is naturally defined as $\mathrm{DART} := P^{\rm DA} - P^{\rm RT}$, and virtual trading positions are defined as follows:
    \begin{enumerate}[label=$(\roman*)$]\itemsep -2pt
        \item An {\rm \bf INC (incremental)} bid is a virtual supply position in which the trader \emph{sells} a quantity $q > 0$ of energy $($in MWh$)$ in the Day-Ahead Market and later \emph{buys} it back in the Real-Time Market, yielding a payoff of $q \cdot \mathrm{DART}$. {\rm INC} positions are profitable when the Day-Ahead Market overestimates real-time prices.
        \item A {\rm \bf DEC (decremental)} bid is a virtual demand position in which the trader \emph{buys} a quantity $q > 0$ of energy $($in MWh$)$ in the Day-Ahead Market and later \emph{sells} it back in the Real-Time Market, yielding a payoff of $-q \cdot \mathrm{DART}$. {\rm DEC} positions are profitable when the Day-Ahead Market underestimates real-time prices.
    \end{enumerate}
\end{definition}}

\revision{Since INC and DEC positions allow market participants to profit from DART spreads in either direction, the problem faced by a virtual trader is both predictive and operational:
\begin{enumerate}
    \item \textbf{Forecasting:} identify when and where large DART spreads are likely to occur, by exploiting predictable patterns in congestion, load forecasts, and market conditions available at the time of day-ahead bidding;
    \item \textbf{Optimal execution:} translate these directional signals into optimal day-ahead positions, determining not only whether to trade but also how much to trade in each zone, while accounting for the endogenous feedback between submitted quantities and day-ahead clearing prices.
\end{enumerate}
This dual nature motivates the joint framework developed in this paper, which combines a spike-forecasting model with a structural price impact model to determine both the direction and the size of virtual positions across zones.}

A growing literature studies DART spreads and virtual bidding from both predictive and structural perspectives, emphasizing the role of congestion, risk premia, and limits to arbitrage in two-settlement electricity markets. Early empirical work documents the prevalence and economic drivers of DART price deviations and the role of congestion and forecast errors in shaping these spreads \cite{borenstein2002measuring}. \revision{Price dynamics in electricity markets are also shaped by spillovers from related commodity markets such as natural gas, coal, and carbon emissions \cite{ioannidis2025spillovers}.} From an empirical standpoint, extreme DART events are documented to be short-lived, clustered, and closely linked to binding network constraints and unexpected demand shocks \cite{christensen2012forecasting, liebl2013modeling, sandhu2016electricity, sandhu2016ontario, deschatre2020estimating}. This clustering has in particular motivated self-exciting point-process (Hawkes-type) models for spike occurrences in electricity prices \cite{christensen2009never,herrera2014modeling,clements2015modelling,eyjolfsson2018self,deschatre2022electricity}. 
\revision{More broadly, the statistical modelling of electricity prices has attracted significant attention, with approaches ranging from fundamental equilibrium models arising from the intersection of aggregate supply and demand curves, with local properties of the bid stack governing price sensitivity to quantity shocks \cite{howison2009stochastic}, to reduced-form time-series specifications \cite{ioannidis2021electricity}.} 
More recently, machine-learning approaches have been applied to electricity price and DART forecasting, showing that extreme price dislocations can be predicted with economically meaningful accuracy \cite{lago2021forecasting,galarneau2022foreseeing,wang2024deeplearningbasedelectricityprice,forgetta2025dart}. 
\revision{Among these, \cite{galarneau2022foreseeing, wang2024deeplearningbasedelectricityprice, forgetta2025dart} also examine the profitability and limits of virtual bidding strategies, emphasizing the importance of transaction costs, market power, and convergence effects.}
Other more structural approaches that jointly model load and price dynamics for risk management and hedging in electricity markets include \cite{coulon2013model,ernstsen2017hedging}, and complementary model-based formulations of strategic bidding and intraday trading can be found in \cite{aid2016optimal,morri2024learning}.

Among the studies mentioned above, \cite{galarneau2022foreseeing} provides a predictive framework for identifying and trading extreme DART spikes in Long Island, the second largest zone in NYISO, and demonstrates that spike forecasting is feasible and economically valuable. \revision{However, their study restricts trades to a fixed unit size of 1 MWh, which renders price impact negligible by construction and limits the practical relevance of the strategy for large traders. In practice, a virtual trader optimizes the submitted quantity to maximize expected profits, but submitting substantially larger quantities mechanically shifts the day-ahead clearing price, eroding the very spreads they seek to exploit. A principled model of price impact is therefore essential to translate predictive signals into economically meaningful positions at scale. Beyond the price impact issue, their framework is also restricted to a single zone, whereas in markets such as NYISO, where DART spreads exhibit substantial cross-zonal heterogeneity driven by localized congestion and transmission constraints, a multi-zone approach can capture additional diversification benefits and allow risk to be concentrated where marginal price impact is lowest. Overall, determining how much to trade in each zone, while accounting for cross-zone interactions and heterogeneous price impacts, is essential to maximizing the economic value of the predictive signals. This paper addresses these challenges by developing a unified framework that jointly optimizes forecasting and position sizing across zones.}

More precisely, this paper extends the framework of \cite{galarneau2022foreseeing} in four key directions:
\begin{enumerate}[itemsep=0em]
    \item \textbf{Multi-zone, two-sided DART spike forecasting.} We jointly model extreme positive and negative DART spreads across all NYISO zones, allowing for correlated bi-directional spike dynamics across locations.
    \item \textbf{A structural, economically consistent model of market impact.} \revision{Correctly sizing trades requires an explicit model of price impact, as large virtual positions mechanically shift the day-ahead clearing price. Using day-ahead bid stacks, we estimate system-wide energy impact coefficients, capturing how net long or short virtual load shifts the day-ahead clearing price. Zone-specific congestion sensitivities are calibrated from zonal load and price data. Together,} this yields a linear-quadratic impact model linking trade size to expected price perturbation.
    \item \textbf{Optimal scaling of virtual positions.} We derive closed-form expressions for the profit-maximizing vector of zonal quantities, incorporating asymmetric buy/sell impacts and cross-zone interactions. This allows us to determine how large a trade should be in each zone, not merely whether a trade should be executed. 
    \item \textbf{Empirical validation at scale.}  When deployed on 2022--2025 out-of-sample data in multiple zones of three ISO regions, NYISO, ISO--NE and ERCOT, the resulting strategy achieves substantial profitability and remains robust across market regimes, including the extreme heat-wave events observed during Summer~2025.
\end{enumerate}

Our methodology proceeds in three stages. First, we train in Section~\ref{sec:empirical-multi-iso} zone-level classifiers to forecast extreme positive and negative DART events using historical load, price, and congestion features described in Section~\ref{sec:model_setup}. \revision{Second, in Section~\ref{sec:scaling-impact}, we develop a structural price impact model that yields closed-form expressions for the optimal vector of zonal virtual positions, explicitly accounting for asymmetric system-wide and local market impact (Section~\ref{sec:optimal}). System-wide energy impact coefficients are estimated from the local slope of the day-ahead bid stack (Section~\ref{sec:modelling_ke}), while zone-specific congestion sensitivities are calibrated from zonal load and price data (Section~\ref{sec:estimating-kz}). This separation between signal generation and impact-aware sizing allows predictive accuracy and economic consistency to be evaluated independently. The resulting strategy is evaluated out-of-sample on NYISO data from 2022--2025 in Section~\ref{sec:performance}.}

By combining predictive modeling, structural analysis of bid stacks, and multi-zone optimization, this paper provides a comprehensive framework for scalable and economically consistent virtual trading in U.S. wholesale electricity markets. Our results show that DART forecasting and position sizing must be treated as a joint problem: forecasts alone are insufficient unless paired with a principled model of price impact and a rigorous scaling rule, as ignoring this feedback either overstates achievable profits or implicitly assumes unrealistically small position sizes. \revision{Empirically, the framework delivers substantial out-of-sample profitability when deployed across multiple zones and markets, with profits concentrated in periods of extreme market dislocations. This pattern is less suggestive of a stable mechanical arbitrage and more consistent with compensation for bearing tail risk during stressed system conditions. Whether this reflects a risk premium, systematic forecast errors, or a combination of both ultimately remains an open question, as resolving it would require a formal comparison against an appropriate risk benchmark, which we leave for future research.}

\section{Predictive Framework and Statistical Model}\label{sec:model_setup}

\begin{figure}[H]
    \centering
    \begin{subfigure}[b]{0.33\textwidth}
        \centering
        \includegraphics[width=\linewidth]{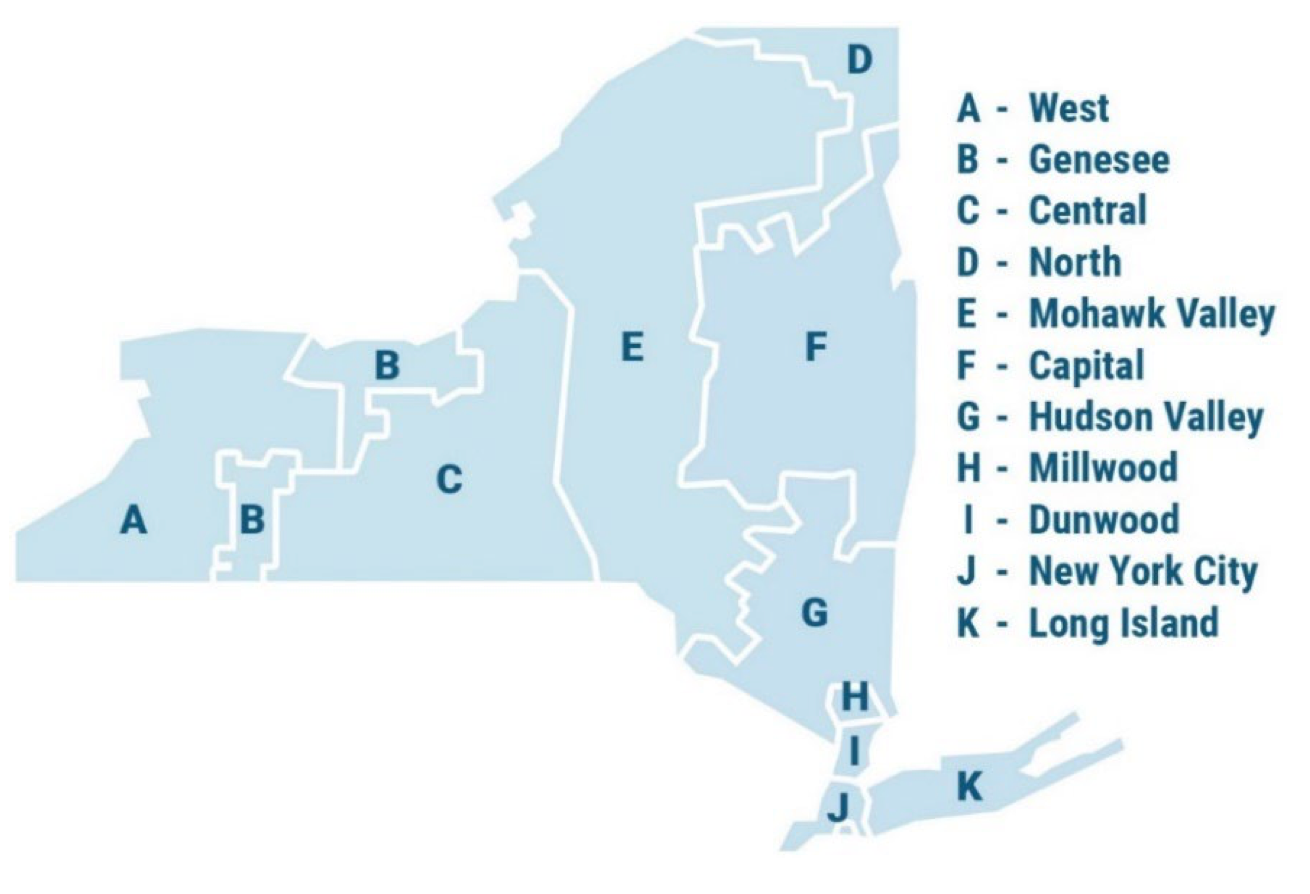}
        \caption{NYISO load zones.}
    \end{subfigure}
    \hfill
    \begin{subfigure}[b]{0.32\textwidth}
        \centering
        \includegraphics[width=\linewidth]{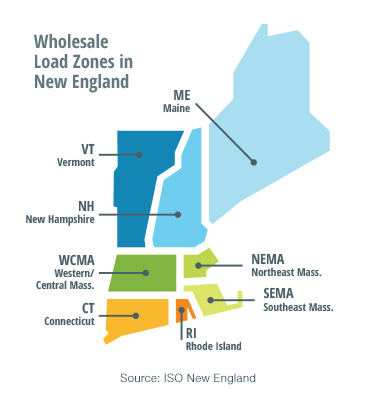}
        \caption{ISO--NE load zones.}
    \end{subfigure}
    \hfill
    \begin{subfigure}[b]{0.33\textwidth}
        \centering
        \includegraphics[width=\linewidth]{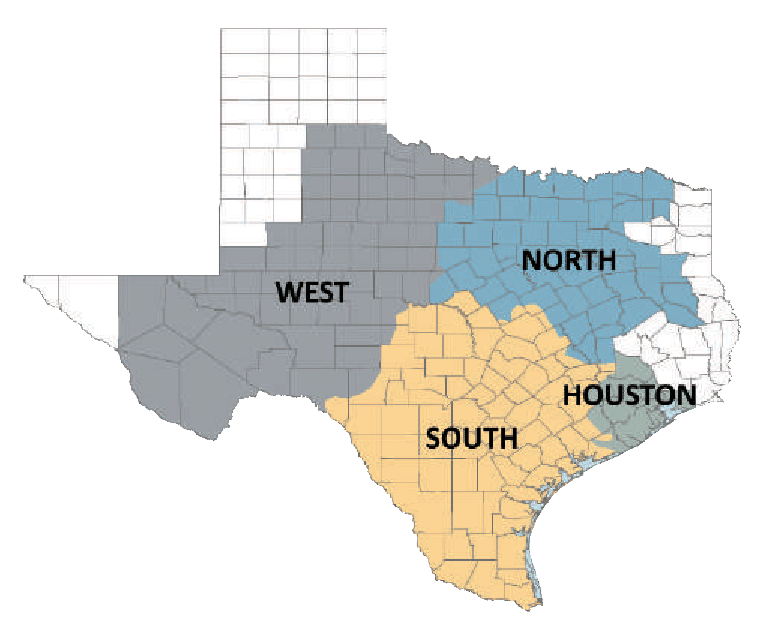}
        \caption{ERCOT load zones.}
    \end{subfigure}
    \caption{Zonal maps for NYISO, ISO--NE, and ERCOT.}
    \label{fig:all-zones-maps}
\end{figure}

This section develops a unified framework for forecasting day-ahead versus real-time (DART) price spreads in U.S. wholesale electricity markets, with an empirical focus on NYISO, ISO--NE, and ERCOT. Figure~\ref{fig:all-zones-maps} shows the zonal layouts for the three regions.

\subsection{Data Construction}

For each of the three markets, we construct an hourly panel containing day-ahead and real-time prices, system and zonal load forecasts, and a set of exogenous covariates. These include lagged DART values (24h/48h), lagged load forecast errors, hour-of-day and month-of-year indicators, holiday/weekend dummies, and seasonal effects. In practice, all features are constructed using a common set of definitions, lag structures, and calendar conventions across the markets, enabling direct comparability. All market data were obtained from the public data portals of the corresponding system operators, namely ERCOT \cite{ercot_data}, ISO New England \cite{isone_data}, and NYISO \cite{nyiso_data}. 

Data access and aggregation were facilitated using the GridStatus platform (\url{https://www.gridstatus.io})\revision{, which compiles publicly available data from U.S. system operators, including ISO New England (\url{https://www.iso-ne.com}), the New York ISO (\url{https://www.nyiso.com}), and ERCOT (\url{https://www.ercot.com})}. 
\revision{These variables correspond to the standard information set available to market participants at the time of day-ahead bidding and capture key economic drivers of DART spreads, including congestion dynamics, demand uncertainty, and system conditions. A concise summary of sample periods and modeling choices across the three ISOs is provided in Table~\ref{tab:iso_summary} of Appendix~\ref{app:data_summary}.}

\subsection{Feature Labeling}

As day-ahead bids must be submitted well in advance of the operating day, all predictors are constructed using only data available strictly before the corresponding Day-Ahead Market’s gate-closure time, \revision{reported in Table~\ref{tab:horizons} below. The prediction horizon is defined as the time elapsed between gate closure and each hourly delivery period (00:00--23:00) of the following operating day, and therefore varies across both markets and hours. This ensures that the forecasting exercise is entirely free of look-ahead bias and that the timing of the prediction problem is transparent and consistent across markets.}

\begin{table}[h]
\centering
\begin{tabular}{lccc}
\toprule
Market & Gate-closure & Prediction horizon \\
\midrule
NYISO   & 05:00 & 19--42h \\
ISO--NE & 10:30 & 13.5--36.5h \\
ERCOT   & 10:00 & 14--37h \\
\bottomrule
\end{tabular}
\caption{Day-Ahead Market gate-closure times and corresponding prediction horizons.}
\label{tab:horizons}
\end{table}

\revision{The feature vector $X_{t,z,m} \in \mathbb R^d$ available at the day-ahead decision time for the operating hour $t$, in zone $z$ of market $m$, is constructed using only information available prior to the Day-Ahead Market close and consists of 
\begin{enumerate}[label=$(\roman*)$]\itemsep -2pt
    \item \textit{Lagged {\rm DART} values at 24h and 48h horizons}: transmission constraints and system conditions affecting DART spreads tend to cluster over time, so that previous DART realizations help capture the persistence of congestion regimes and short-term market imbalances. The 24h and 48h lag structure reflects the natural periodicity of electricity markets: the 24h lag captures the same hour on the previous operating day, which is the most relevant reference point for day-ahead participants, while the 48h lag provides an additional anchor that helps distinguish transient imbalances from more persistent congestion patterns.
    \item \textit{Zonal and system-level day-ahead load forecasts}: higher anticipated demand increases system stress and the likelihood of binding transmission constraints, making load forecasts a natural proxy for expected congestion and the anticipated level of price divergence between day-ahead and real-time settlements.
    \item \textit{Lagged zonal and system-level load forecast errors}: systematic or volatile deviations between forecasted and realized demand are a key driver of DART spreads; recent forecast errors therefore serve as indicators of forecasting bias and demand uncertainty in the day-ahead market.
    \item \textit{Calendar indicators for hour-of-day, month-of-year, and holidays/weekends}: electricity demand, generation mix, and operational constraints follow predictable intraday and seasonal patterns, giving rise to systematic differences in congestion and price formation across peak and off-peak periods.
    \item \textit{Season-of-year indicators $($Winter, Summer, Shoulder$)$}: broader structural conditions such as summer peak demand or winter fuel constraints generate distinct regimes of price volatility and spike behavior, which seasonal indicators help the model capture.
\end{enumerate}}

The dimension $d$ corresponds to the total number of covariates after feature construction, 
and depends on the market and feature availability. For example, in the NYISO application considered below, this results in a feature dimension of $d = 50$, significantly larger than $d = 9$ used in \cite{galarneau2022foreseeing}. Specifically, each hourly observation is represented by a 50-dimensional feature vector $X_t$, constructed by concatenating four zone-level predictors---day-ahead load forecasts, lagged DART values (24h and 48h), and lagged load forecast errors---for each of the 11 zones ($4 \times 11 = 44$), together with six calendar covariates encoding weekend, holiday, diurnal, and seasonal effects. \revision{Crucially, by including these predictors for all zones simultaneously, the feature vector captures cross-zonal heterogeneity in congestion exposure and network topology without requiring explicit zone fixed effects. Zone-specific predictive relationships are then learned separately for each location through the logistic regression models described in the next section, via coefficients later denoted $\beta_{z,m}$.}

\subsection{Spike Definition and Logistic Regression Models}

Given the DART value at time $t$, in zone $z$ of market $m$, we define binary labels for negative and positive DART spikes as
\[
y^{\mathrm{neg}}_{t,z,m}
=
\mathbf{1}_{\{\mathrm{DART}_{t,z,m} \le -\gamma_{\mathrm{neg}}(m)\}},
\qquad
y^{\mathrm{pos}}_{t,z,m}
=
\mathbf{1}_{\{\mathrm{DART}_{t,z,m} \ge \gamma_{\mathrm{pos}}(m)\}},
\]
where the market-specific thresholds $\gamma_{\mathrm{neg}}(m)$ and $\gamma_{\mathrm{pos}}(m)$ are calibrated through exploratory analysis and validation. These labels isolate the economically meaningful extremes of the DART distribution that are most relevant for virtual trading strategies. 

Our first objective is to perform a logistic regression to predict DART spikes. For each zone $z$ of market $m$ and spike type, we define the predicted spike probability at time $t$ as
\begin{align}\label{eq:predic_prob}
    p_{t,z,m} = \mathbb{P}(y_{t,z,m}=1 \mid X_{t,z,m})
= \sigma(\beta_{z,m}^\top X_{t,z,m}),
\quad \text{with} \quad
\sigma(u)=\frac{1}{1+e^{-u}}.
\end{align}
The coefficients $\beta_{z,m}$ introduced in the above equation are obtained by minimizing the following cross-entropy loss for each zone $z$ of market $m$:
\[
\min_{\beta}
\sum_{t}
 {\Big[
 y_{t,z,m}\left(-\log p_{t,z,m}\right)
+ (1-y_{t,z,m})\left(-\log(1-p_{t,z,m})\right)
\Big]}.
\]
The training windows for the logistic regression differ across markets due to data availability (NYISO: 2015--2019; ISO--NE and ERCOT: 2018--2022), and a separate validation period is then used to tune probability thresholds.

\medskip

\noindent\textbf{Model selection.} 
Before settling on logistic regression, we performed an extensive comparison across several supervised learning methods, including random forests, gradient-boosted trees, feed-forward neural networks \cite{hastie2009elements}, \revision{and SMOTE-based resampling to address the rarity of spike events in the training data.} While some nonlinear models achieved marginally higher in-sample classification accuracy, none offered a consistent improvement in out-of-sample trading P\&L. In particular, spike events are extremely rare and more flexible models tend to overfit the noise in the training data, producing unstable probability forecasts that translate poorly into trading signals. \revision{In the end, logistic regression delivered the most robust---and interpretable---out-of-sample performance across all three markets. This conclusion is consistent with \cite{galarneau2022foreseeing}, who report analogous results for the Long Island zone in NYISO and find that logistic regression performs best among the methods they consider when evaluated on economic grounds. We therefore do not present an exhaustive quantitative benchmarking exercise here, since our aim is not to optimize forecast accuracy per se, but to identify reliable and economically relevant trading signals.}

\section{Empirical {Performance of Benchmark Strategies}}
\label{sec:empirical-multi-iso}
In this section, we study DART spreads and the performance of the benchmark spike-based INC/DEC strategies described below in Section \ref{benchstrats}, across the three major U.S. power markets: NYISO, ISO--NE, and ERCOT. All analyses use hourly data and a common modeling framework, with separate classifiers for positive and negative DART spikes calibrated on a validation set and evaluated out-of-sample.

We emphasize that pooling information across zones is essential in NYISO, where congestion and losses generate substantial zonal heterogeneity. In contrast, ISO--NE and ERCOT exhibit highly synchronized zonal DART movements, so that a single representative zone captures most of the relevant variations.

\subsection{{Benchmark INC and DEC Strategies}}\label{benchstrats}

Recall that $p_{t,z,m}$ denotes the predicted probability of a DART spike at time $t$ in zone $z$ of market $m$, as defined in \eqref{eq:predic_prob}. We follow the approach in \cite{galarneau2022foreseeing}, assuming that a trade is executed whenever
\[
p_{t,z,m} \ge \tau_{z,m},
\]
where $\tau_{z,m}$ is a zone-specific threshold. The predictive model used to estimate the probability \(p_{t,z,m}\) of a DART spike, conditional on the feature vector \(X_{t,z,m}\), is trained on a historical training set, while $\tau_{z,m}$, $\gamma_{\text{neg}}(m)$ and $\gamma_{\text{pos}}(m)$ are selected to maximize P\&L on a separate validation set under unit-size trading and no price impact. All performance results reported below are evaluated on a held-out test set.

We study two benchmark strategies:\vspace{-0.5em}
\begin{enumerate}[label=$(\roman*)$,itemsep=0.5pt]
    \item \textbf{INC-only}: trade only when a positive DART spike is predicted, earning $+\mathrm{DART}_{t,z,m}$;
    \item \textbf{DEC-only}: trade only when a negative DART spike is predicted, earning $-\mathrm{DART}_{t,z,m}$.
\end{enumerate}\vspace{-0.5em}
These strategies provide a clean baseline for comparing predictive performance
across markets and zones. Building on this benchmark framework, we will extend in Section~\ref{sec:optimal} the
single-zone trading rule to a joint multi-zone optimization problem with
endogenous position sizing and price impact.

\subsection{P\&L in NYISO }\label{sec:nyiso}
For NYISO, we work with hourly data from 2015--2025 across eleven load zones, and
focus the discussion on six large-demand zones: CAPITL, CENTRL, LONGIL, NORTH,
NYC, and WEST. We adopt the following chronological split
\[
\text{Train: } 2015\text{--}2019,\qquad
\text{Validation: } 2020\text{--}2021,\qquad
\text{Test: } 2022\text{--}2025.
\]
\revision{This validation-based procedure also provides robustness to the choice of spike thresholds, feature specification and training window.}
Separate logistic classifiers are fit for positive and negative DART spikes on
the training set, with spike thresholds and probability cutoffs selected on the
validation set to maximize unit-size P\&L. The resulting thresholds used in the
NYISO analysis are
\[
\gamma_{\text{pos}} = 5\$/\text{MWh},\qquad
\gamma_{\text{neg}} = 30\$/\text{MWh}.
\]
We tune the probability cutoffs $\tau_{\text{pos}}$ and $\tau_{\text{neg}}$
separately for each zone. For example, in NYC the selected cutoffs are
$(\tau_{\text{pos}}, \tau_{\text{neg}}) = (0.75, 0.9)$, while for Long Island
they are $(0.7, 0.9)$.

\begin{figure}[H]
  \centering

  \begin{subfigure}[t]{0.49\textwidth}
    \centering
    \includegraphics[width=\linewidth,height=0.13\textheight,keepaspectratio]{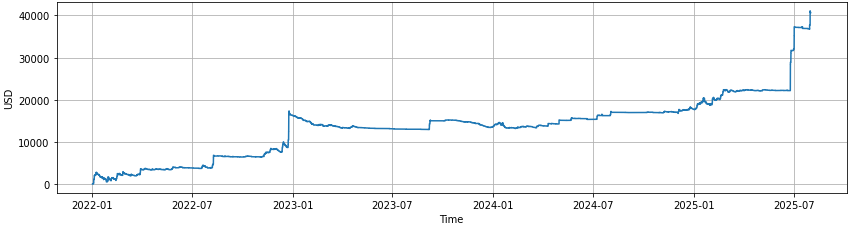}
    \caption{NYC — DEC}
  \end{subfigure}\hfill
  \begin{subfigure}[t]{0.49\textwidth}
    \centering
    \includegraphics[width=\linewidth,height=0.13\textheight,keepaspectratio]{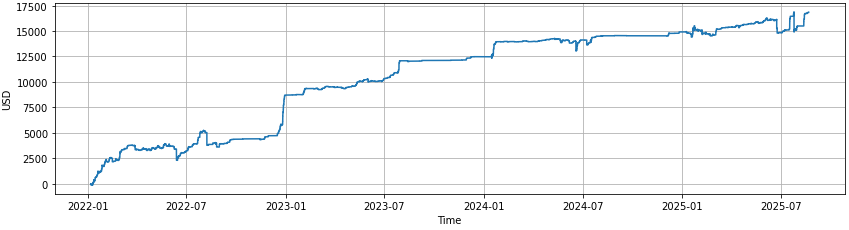}
    \caption{NYC — INC}
  \end{subfigure}

  \vspace{0.6em}

  \begin{subfigure}[t]{0.49\textwidth}
    \centering
    \includegraphics[width=\linewidth,height=0.13\textheight,keepaspectratio]{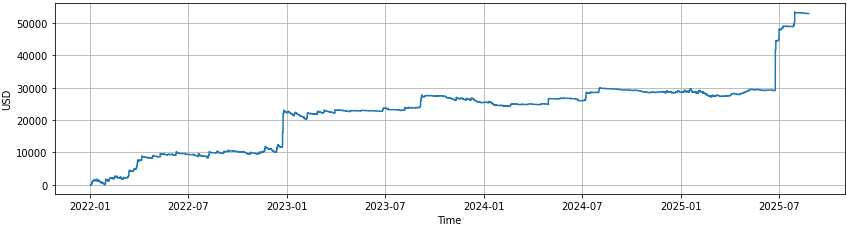}
    \caption{LONGIL — DEC}
  \end{subfigure}\hfill
  \begin{subfigure}[t]{0.49\textwidth}
    \centering
    \includegraphics[width=\linewidth,height=0.13\textheight,keepaspectratio]{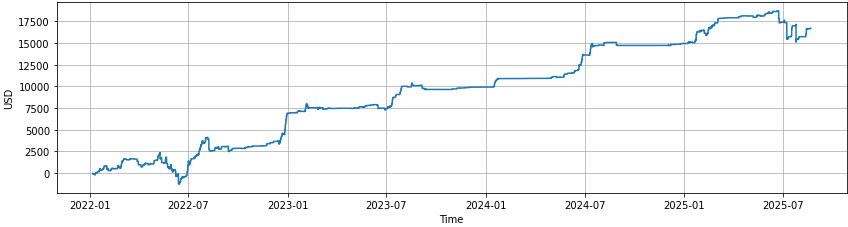}
    \caption{LONGIL — INC}
  \end{subfigure}

  \caption{NYISO: cumulative P\&L for NYC and Long Island zones under the DEC/INC benchmark strategy.}
  \label{fig:zone-pnls-main}
\end{figure}

Figures~\ref{fig:zone-pnls-main}(a)--(d) report cumulative P\&L over the
2022--2025 test period for INC-only and DEC-only benchmark strategies in NYC and Long Island, the two zones with the highest demand. Corresponding results for the remaining zones are presented in Figures~\ref{fig:zone-pnls-appendix}(a)--(h) in the Appendix. These figures highlight pronounced cross-zonal heterogeneity: while some regions exhibit persistent profitability from DEC positions, others display stronger performance for INC trading.

Tables~\ref{tab:zone-pnls-inc} and~\ref{tab:zone-pnls-dec} (in the Appendix) report cumulative P\&L by zone, which we relate to the yearly mean DART spreads shown in Table~\ref{tab:dart-means}. \revision{Moreover, these tables also report annualized volatility (in USD, consistent with the P\&L-based formulation) and Sharpe ratios, allowing for a comparison of risk-adjusted performance across zones. }Zones with systematically positive DART averages (e.g., CAPITL or, in earlier years, LONGIL) tend to favour INC strategies, whereas zones with negative or mixed averages (such as NYC in the post-2022 period) are more aligned with DEC trading. This structural bias in the DART distribution helps explain cross-zonal differences in profitability and provides guidance on whether a zone is better approached with INC-only, DEC-only, or mixed strategies.

Cross-zone dependence remains strong throughout NYISO, as shown in Table~\ref{tab:dart-corr}, but varies meaningfully across groups of zones. Upstate zones exhibit particularly high correlations in DART spreads, while downstate zones form a tightly interconnected cluster. This pattern reflects localized congestion and loss effects superimposed on system-wide price movements, and motivates pooling information across zones while retaining zone-specific features in the predictive models. 

\subsection{P\&L in ISO--NE}\label{sec:isone}

We perform a parallel analysis on ISO New England, which consists of eight load zones, using hourly data from November 2018 to October 2025. DART spreads across ISO--NE zones are almost perfectly correlated, as shown in Table~\ref{tab:isone_dart_corr}, indicating that economically relevant variation is predominantly system-wide rather than zonal. To avoid redundant signals, we therefore conduct the spike-prediction and trading analysis on a single representative zone, Maine (ME).

To account for a shorter data sample compared to NYISO, we adopt the following split
\[
\text{Train: } 2018\text{--}2022,\qquad
\text{Validation: } 2023,\qquad
\text{Test: } 2024\text{--}2025.
\]
Separate classifiers are fit for positive and negative DART spikes, with spike thresholds and probability cutoffs tuned on the validation set. In particular, for the Maine load zone in ISO--NE, the resulting parameters are
\[
\gamma_{\text{pos}} = 2\$/\text{MWh},\qquad
\gamma_{\text{neg}} = 8\$/\text{MWh},\qquad
\tau_{\text{pos}} = 0.70,\qquad
\tau_{\text{neg}} = 0.90.
\]

\begin{figure}[ht!]
    \centering
    \begin{subfigure}[t]{0.32\textwidth}
        \centering
        \includegraphics[width=\linewidth,height=0.15\textheight,keepaspectratio]{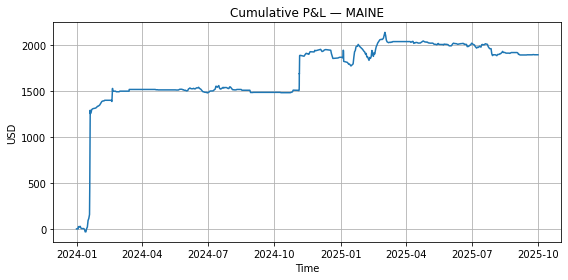}
        \caption{Total P\&L (ME).}
    \end{subfigure}
    \hfill
    \begin{subfigure}[t]{0.32\textwidth}
        \centering
        \includegraphics[width=\linewidth,height=0.15\textheight,keepaspectratio]{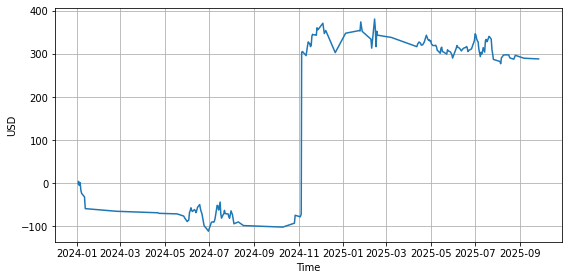}
        \caption{DEC-only P\&L.}
    \end{subfigure}
    \hfill
    \begin{subfigure}[t]{0.32\textwidth}
        \centering
        \includegraphics[width=\linewidth,height=0.15\textheight,keepaspectratio]{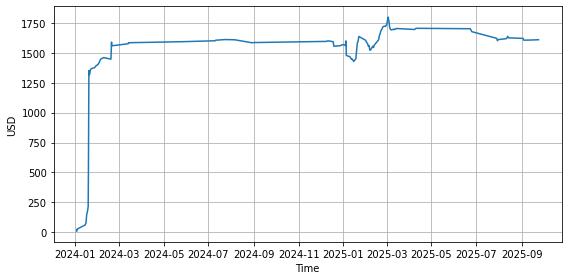}
        \caption{INC-only P\&L.}
    \end{subfigure}
    \caption{ISO--NE Maine (ME) zone: cumulative P\&L curves for overall, INC-only, and DEC-only strategies on the 2024--2025 test period.}
    \label{fig:maine-three-panel-isone}
\end{figure}

Figure~\ref{fig:maine-three-panel-isone} {shows the cumulative P\&L curves for the INC-only and DEC-only benchmark strategies} on the 2024--2025 test period. Overall performance is weaker than in NYISO, with most profits arising from INC trades, while DEC positions are triggered less frequently due to both the smaller negative-spike threshold and the model’s lower predictive sharpness in this market.

\subsection{P\&L in ERCOT}\label{sec:ercot}

Finally, we analyze the ERCOT market using hourly data from 2018--2025 across its four principal load zones: North, South, West, and Houston. Similarly to ISO--NE, DART spreads across ERCOT zones are extremely highly correlated (see Table~\ref{tab:ercot-dart-corr}), with pairwise correlations exceeding $0.97$, indicating that DA–RT dynamics are effectively system-wide. As a result, all zones generate nearly identical predictions and trading behavior. We therefore restrict attention on the \textsc{WEST} zone for concreteness.

\begin{figure}[ht!]
\centering
\begin{subfigure}[t]{0.32\textwidth}
  \centering
  \includegraphics[width=\linewidth]{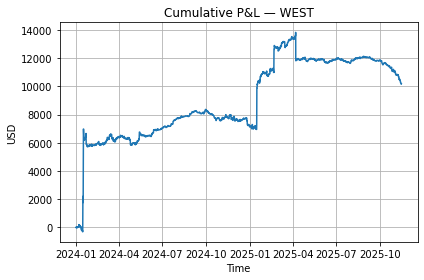}
  \caption{Total P\&L (WEST).}
\end{subfigure}\hfill
\begin{subfigure}[t]{0.32\textwidth}
  \centering
  \includegraphics[width=\linewidth]{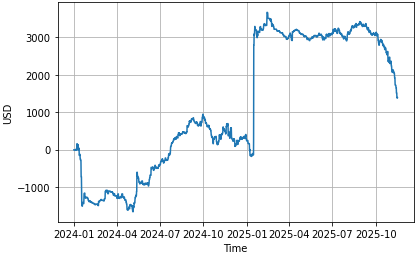}
  \caption{DEC-only P\&L.}
\end{subfigure}\hfill
\begin{subfigure}[t]{0.32\textwidth}
  \centering
  \includegraphics[width=\linewidth]{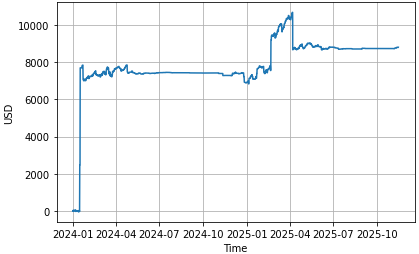}
  \caption{INC-only P\&L.}
\end{subfigure}
\caption{ERCOT WEST zone: cumulative P\&L curves for overall, INC-only, and DEC-only strategies on the 2024--2025 test period.}
\label{fig:west-three-panel}
\end{figure}

To account for a shorter data sample compared to NYISO, we adopt the following split
\[
\text{Train: } 2018\text{--}2022,\qquad
\text{Validation: } 2023,\qquad
\text{Test: } 2024\text{--}2025.
\]
As before, spike thresholds and probability cutoffs are tuned on the validation set. For WEST, the resulting parameters are
\[
\gamma_{\text{pos}} = 15\$/\text{MWh},\qquad 
\gamma_{\text{neg}} = 10\$/\text{MWh},\qquad 
\tau_{\text{pos}} = 0.75,\qquad 
\tau_{\text{neg}} = 0.90 .
\]
Figure~\ref{fig:west-three-panel} reports the total, INC-only, and DEC-only benchmark strategy P\&L curves on the 2024--2025 test period. The strategy is predominantly driven by INC trades, reflecting the systematic tendency of the ERCOT day-ahead market to overestimate real-time prices during this period, with most gains concentrated in a sharp episode early in 2024. DEC positions, which profit from negative DART spikes, are initially loss-making before recovering, with a concentrated burst of profits at the beginning of 2025, after which performance stabilizes at a modest positive level.

\subsection{Cross–Market Comparison}\label{sec:cross_market}

To provide distributional context for the spike thresholds and trading activity across markets, Table~\ref{tab:dart-quantiles-longil-maine-west} reports empirical DART quantiles on the training samples for the following three representative zones: NYISO LONGIL, ISO--NE ME, and ERCOT WEST. NYISO exhibits heavier intermediate and upper tails, with substantially larger $90$th–$99$th percentiles relative to ISO--NE and ERCOT. Taken together, the heavier-tailed DART distribution in NYISO helps explain its greater trading profitability relative to ISO--NE and ERCOT. More precisely, larger and more frequent spikes generate a richer set of economically meaningful opportunities, on which predictive signals can be exploited more often. Differences in observed INC and DEC activity across markets therefore reflect both underlying market structure and the predictive sharpness of the model, rather than spike magnitudes alone.

Tables~\ref{tab:dart-corr}, \ref{tab:isone_dart_corr}, and \ref{tab:ercot-dart-corr} highlight pronounced differences in cross-zonal DART dependence across markets. More precisely, in ERCOT and ISO--NE markets, DART spreads are almost perfectly synchronized across zones, indicating that deviations between day-ahead and real-time prices are driven primarily by system-wide factors. \revision{Therefore, in these markets, the benefits of a detailed multi-zone structural model are more limited, and the analysis of a single representative zone accounts for most of the relevant variations.} By contrast, the NYISO market exhibits substantially weaker and more heterogeneous cross-zonal correlations, reflecting localized congestion, losses, and transmission constraints. \revision{This structural heterogeneity implies that DART dynamics in NYISO cannot be reduced to a single system factor, making multi-zone modelling essential.} 

\revision{In the following section, we develop a general optimal trading framework, theoretically applicable across all ISO markets. Nevertheless, its incremental value is greatest in markets exhibiting substantial cross-zonal heterogeneity. As extending the single-zone framework of \cite{galarneau2022foreseeing} to a multi-zone setting is precisely one of the main contributions of this paper, we therefore focus its empirical implementation on the NYISO market, where localized congestion effects generate richer and more diverse trading opportunities across zones.}

\section{Optimal Trading Strategy}\label{sec:scaling-impact}

A central objective of this section is to determine how large our virtual trading positions should be across zones, once a directional signal has been generated by the spike-forecasting model. Correctly sizing trades is essential: although DART spreads create strong economic opportunities, large virtual positions mechanically shift the day-ahead clearing price through both system-level and zonal congestion effects. Hence, to maximize profitability while avoiding excessive market impact, we require an explicit model linking trade size to day-ahead price response. \revision{The following subsections present the price impact model, discuss the estimation of its parameters, and derive the resulting optimal zonal quantities. The framework developed in this section is general and theoretically applicable across ISO markets. However, as argued in Section~\ref{sec:cross_market}, its incremental value is greatest in markets with substantial cross-zonal heterogeneity; we therefore focus its empirical implementation on NYISO, where localized congestion effects make multi-zone modelling particularly informative. For notational simplicity, we henceforth drop the market subscript $m$.}

\subsection{Price Impact {Model}}
Let $Z$ be the number of zones in the considered market, and let $q_t \in \mathbb{R}^Z$ be the vector of signed bidding quantities (MWh), where $q_{t,z}>0$ represents an INC (virtual supply) in zone $z$ at time $t$ and  $q_{t,z}<0$ represents a DEC (virtual demand). For each hour $t$ and zone $z$, denote
\[
\text{DA}_{t,z},\;\text{RT}_{t,z} \in \mathbb{R},
\qquad
\text{DART}_{t,z} := \text{DA}_{t,z}-\text{RT}_{t,z}.
\]
The trading edge (in \$/MWh) of an INC or a DEC trade is then
\[
r^{\text{INC}}_{t,z} = \text{DART}_{t,z},
\qquad
r^{\text{DEC}}_{t,z} = - \text{DART}_{t,z}.
\]
If a trade of size $q_t \in \mathbb R^Z$ is executed at time $t$, the realized dollar P\&L for side $s\in\{\text{INC},\text{DEC}\}$ in each zone $z$ is
\begin{equation}\label{eq:pnl-basic}
    \Pi^{(s)}_{t,z}(q_t)
    \;=\;
    q_{t,z} \Bigl( r^{(s)}_{t,z} - I_{t,z}(q_t) \Bigr),
\end{equation}
where $I$ is the price impact function (in \$/MWh) imposed on the DA price, depending on the submitted quantity, time and zone. 

Before specifying the price–impact model, we recall the standard decomposition of day-ahead locational marginal prices (LMPs). In all U.S. ISOs, including NYISO, ISO--NE, and ERCOT, \revision{day-ahead locational marginal prices follow the standard additive decomposition}
\begin{align}\label{eq:lmp_decomp}
    {\rm DA}_{t,z} = \text{Energy}_t + \text{Loss}_{t,z} + \text{Congestion}_{t,z},
\end{align}
where the energy component is system-wide, whereas losses and congestion vary across zones due to transmission constraints and network topology. These spatial components are precisely what generate cross-zonal heterogeneity in DART spreads and motivate a zone-dependent treatment of price impact in our scaling model. \revision{We emphasize that this decomposition is not a modeling assumption but a fundamental feature of how ISOs clear electricity markets under security-constrained economic dispatch, and directly motivates the additive structure of our impact model, see \cite{nyiso_lbmp_2025, isone_lmp_faq}.}

Virtual demand and supply bids shift the residual demand curve and thus impact day-ahead clearing prices. \revision{To model this impact, we follow the linear framework standard in optimal execution \cite{almgren2001optimal, gatheral2010dynamics}, see also \cite{aid2016optimal} for a stochastic control formulation of optimal trading with price impact in electricity markets. Moreover, given the additive decomposition of day-ahead locational marginal prices in \eqref{eq:lmp_decomp}, we decompose the price impact into a zone-specific component, capturing the local effect of trading in zone $z$, and a system-wide component, capturing the effect of the aggregate net position denoted $S_t$:}
\begin{align}\label{eq:price_impact}
    I_{t,z}(q_t)
= \bigl(k_{\rm E}^{+}\mathbf{1}_{\{S_t\geq0\}} + k_{\rm E}^{-}\mathbf{1}_{\{S_t<0\}}\bigr) S_t + k_z q_{t,z},
\qquad
S_t := \sum_{z = 1}^Z q_{t,z}.
\end{align}

\revision{The first term in \eqref{eq:price_impact} represents the system-wide component, which depends on the sign of the aggregate net position $S_t$. Indeed, when $S_t > 0$, the system takes a net DEC position on the market and the DA price moves along the demand curve, whereas when $S_t < 0$ it takes a net INC position and the price moves along the supply curve. Since the slopes of the supply and demand curves may differ, their marginal price impacts differ as well, motivating the use of distinct energy-impact coefficients $k_{\rm E}^{+}$ and $k_{\rm E}^{-}$. We estimate these coefficients for NYISO in Section~\ref{sec:modelling_ke} using aggregate supply and demand curves.}

\revision{The second term in \eqref{eq:price_impact} captures the local price impact of a position of size $q_{t,z}$ in zone $z$, governed by the zone-specific coefficient $k_z$.} Indeed, zones in NYISO differ substantially in their typical demand levels and exposure to transmission constraints. Large demand centers such as NYC and Long Island tend to absorb incremental virtual positions with relatively small marginal effects on losses and congestion, whereas smaller or more constrained zones can exhibit much larger local price sensitivities. This motivates introducing a zone-specific local impact coefficient $k_z$, capturing how virtual demand or supply submitted  bids in zone $z$ affect the spatial components of day-ahead prices. In Section~\ref{sec:estimating-kz}, we calibrate these coefficients for each zone of the NYISO market using zonal load and price data.

\subsection{Optimal Zonal Quantities}
\label{sec:optimal}

Each zone $z$ is equipped with a classifier that outputs the probability $p_{t,z}$ that hour $t$ in zone $z$ experiences a DEC spike (negative DART) or an INC spike (positive DART). After training on 2015--2019, we tune the decision thresholds $\tau_z$ on the 2020--2021 validation set. At time $t$, we trade in zone $z$ whenever
\[
p_{t,z} \ge \tau_z,
\]
and select the trade direction (INC or DEC) with the larger predicted expected payoff.

Conditional on trading, we next determine the appropriate trade size in each zone. To this end, we estimate the conditional expected economic revenue on the validation period:
\begin{equation}\label{xdef}
x^{\mathrm{INC}}_{t,z} := \mathbb{E}\!\left[r^{\mathrm{INC}}_{t,z} \mid p_{t,z} \ge \tau_z\right],
\qquad
x^{\mathrm{DEC}}_{t,z} := \mathbb{E}\!\left[r^{\mathrm{DEC}}_{t,z} \mid p_{t,z} \ge \tau_z\right].
\end{equation}
\revision{These zone-specific expected payoffs measure the anticipated magnitude of the DART spike in zone $z$ at time $t$, conditional on the model predicting an extreme event. They therefore capture not only whether a trade should be executed, but also how large the expected reward is, which is the key input for determining the optimal position size in each zone.}

Given a fixed time $t$ and the selected trade direction, we collect the expected payoffs into the vector
\[
x_t \equiv (x_{t,1},\dots,x_{t,Z}) \in \mathbb{R}^Z,
\]
and choose zonal quantities $q_{t,z}$ in order to maximize the following objective function, defined for any trade vector $q \in \mathbb{R}^Z$ by
\begin{equation}
F(q)
=\; x_t^\top q
\;-\; \bigl(k_{\rm E}^+\mathbf{1}_{\{S\ge 0\}} + k_{\rm E}^-\mathbf{1}_{\{S<0\}}\bigr) S^2
\;-\; \sum_{z=1}^Z k_z q_z^2,
\qquad S := \mathbf{1}^\top q,
\label{eq:asym-obj}
\end{equation}
which represents the expected price-impacted DART payoff of a trade $q$ placed at time $t$.

We first fix $k_{\rm E}$ (either $k_{\rm E}^{+}$ or $k_{\rm E}^{-}$) and suppose the optimum is interior. The first-order conditions of \eqref{eq:asym-obj} yield the following optimal zonal trade quantities:
\begin{equation}
q_{t,z}^\star = \frac{x_{t,z} - 2k_{\rm E} S}{2k_z}.
\label{eq:qi-star}
\end{equation}
Defining
\[
H \;:=\; \sum_{z=1}^Z \frac{1}{k_z},
\qquad
N_t \;:=\; \sum_{z=1}^Z \frac{x_{t,z}}{k_z},
\]
and summing \eqref{eq:qi-star} over $z$ yields the closed-form net position
\begin{equation}
S = \frac{N_t/2}{1 + k_{\rm E} H}.
\label{eq:S-star}
\end{equation}
We therefore obtain two interior candidate solutions:
\begin{align*}
\text{DEC regime ($S>0$):}\quad
& S^{(+)} = \frac{N_t/2}{1 + k_{\rm E}^{+} H},
&
q_{t,z}^{(+)} = \frac{x_{t,z} - 2k_{\rm E}^{+} S^{(+)}}{2k_z}, \\[0.3em]
\text{INC regime ($S<0$):}\quad
& S^{(-)} = \frac{N_t/2}{1 + k_{\rm E}^{-} H},
&
q_{t,z}^{(-)} = \frac{x_{t,z} - 2k_{\rm E}^{-} S^{(-)}}{2k_z}.
\end{align*}

A third possibility is that the optimum lies on the boundary where the net position is zero, so that the energy impact term vanishes and only local impacts remain. In this regime, we are led to solve
\begin{equation}
\max_{q\in\mathbb{R}^Z}
\; x_t^\top q - \sum_{z=1}^Z k_z q_z^2
\quad\text{subject to}\quad
\mathbf{1}^\top q = 0.
\label{eq:net-flat-problem}
\end{equation}
Introducing a Lagrange multiplier, the optimal zonal trade quantities are given by
\[
q_{t,z}^{(0)} = \frac{x_{t,z} - N_t/H}{2k_z},
\qquad
S^{(0)} = 0.
\]

The optimization therefore yields three candidate solutions $q^{(+)}_t, q^{(-)}_t, q^{(0)}_t$. We retain $q^{(+)}_t$ only when $S^{(+)}>0$, and $q^{(-)}_t$ only when $S^{(-)}<0$. The net-flat solution $q^{(0)}_t$ is always feasible. Among all admissible candidates, the optimal trading vector is the one that maximizes the objective value $F(q)$.

\subsection{Estimating the Energy-impact Coefficients}\label{sec:modelling_ke}

In this section, we develop a local linearization approach to estimate the  system-wide energy-impact coefficients $k_{\rm E}^{+}$ and $k_{\rm E}^{-}$ from the aggregate supply and demand curves observed in the Day-Ahead Market, and calibrate these coefficients on NYISO data. 

\revision{In electricity markets, the Day-Ahead clearing price is determined by the intersection of the aggregate supply and demand curves. More precisely, denoting by $Q^{\rm S}(p)$ and $Q^{\rm D}(p)$ the total quantity supplied and demanded at price $p$, the clearing price $p^\star(t)$ at hour $t$ satisfies}
\[
Q^{\rm S} \bigl(p^\star(t)\bigr) = Q^{\rm D} \bigl(p^\star(t)\bigr).
\]
Equivalently, defining the cleared quantity
\[
q^\star(t) := Q^{\rm S} \bigl(p^\star(t)\bigr) = Q^{\rm D} \bigl(p^\star(t)\bigr),
\]
the relevant price--quantity relationship is given locally by the inverse supply and demand curves $P^{\rm S}(q):=(Q^{\rm S})^{-1}(q)$ and $P^{\rm D}(q):=(Q^{\rm D})^{-1}(q)$ evaluated at $(q^\star(t),p^\star(t))$.

\revision{A net long day-ahead position of size $\Delta q > 0$ corresponds to an exogenous increase in aggregate demand, shifting the residual demand curve upward by $\Delta q$ MWh.} Operationally, the realized cleared quantity becomes
\begin{align}\label{eq:perturbed_price}
    Q^{\rm D}(p^\star) + \Delta q = Q^{\rm D}(p^+) = Q^{\rm S}(p^+) = Q^{\rm S}(p^\star) + \Delta q,
\end{align}
where $p^+$ denotes the perturbed clearing price. \revision{Consistently 
with the linear impact model postulated in \eqref{eq:price_impact}, 
we approximate the price response by a first-order expansion 
of $Q^{\rm S}$ around $p^\star$, which is valid since virtual 
traders submit relatively small quantities compared to total 
cleared volume:
\[
Q^{\rm S}(p^+) \approx Q^{\rm S}(p^\star) + (Q^{\rm S})'(p^\star) 
\cdot (p^+ - p^\star).
\]
This corresponds to a local linearization of the aggregate supply curve around the day-ahead market-clearing point.
Substituting the above approximation in \eqref{eq:perturbed_price} and rearranging yields
\begin{equation}
\label{taylor1}
p^+ - p^\star \approx \frac{\Delta q}{(Q^{\rm S})'(p^\star)} > 0.
\end{equation}
The buy-side price impact is thus governed by the local slope 
of the aggregate supply curve, which defines the energy-impact 
coefficient $k_{\rm E}^{+} = \frac{1}{(Q^{\rm S})'(p^\star)}$. Equivalently, one can write the buy-side price impact using the local slope of the inverse supply curve around $q^\star$, namely $(P^{\rm S})'(q^\star)$, which is precisely the energy-impact coefficient $k_{\rm E}^{+}$.}

\begin{figure}[H]
  \centering
  \begin{subfigure}[t]{0.32\textwidth}
    \centering
    \includegraphics[width=\linewidth]{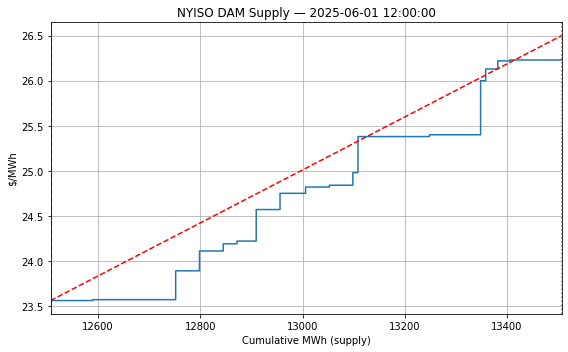}
    \caption{06-01-2025}
    \label{fig:zoom-0501a}
  \end{subfigure}\hfill
  \begin{subfigure}[t]{0.32\textwidth}
    \centering
    \includegraphics[width=\linewidth]{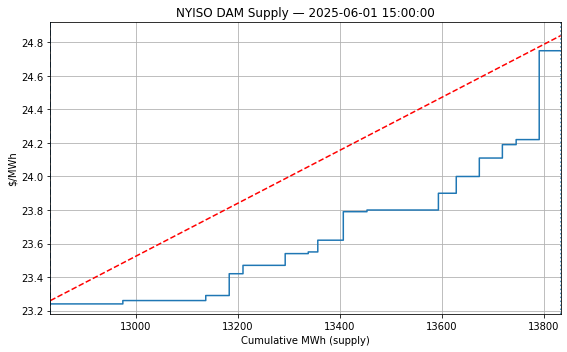}
    \caption{06-01-2025}
    \label{fig:zoom-0301a}
  \end{subfigure}\hfill
  \begin{subfigure}[t]{0.32\textwidth}
    \centering
    \includegraphics[width=\linewidth]{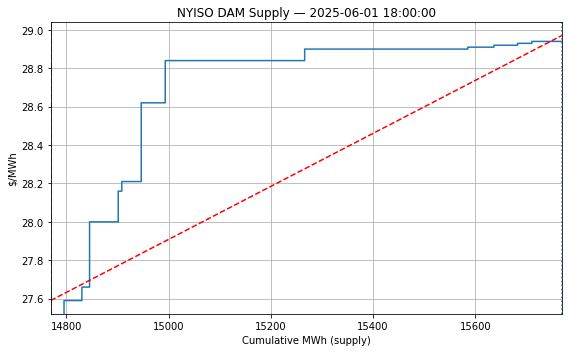}
    \caption{06-01-2025}
    \label{fig:zoom-0701a}
  \end{subfigure}
  \caption{{Supply stack and linear approximation near the DA price-setting intersection point at three different hours}.}
  \label{fig:zoomed-supply-row}
\end{figure}

\begin{figure}[H]
  \centering
  \begin{subfigure}[t]{0.32\textwidth}
    \centering
    \includegraphics[width=\linewidth]{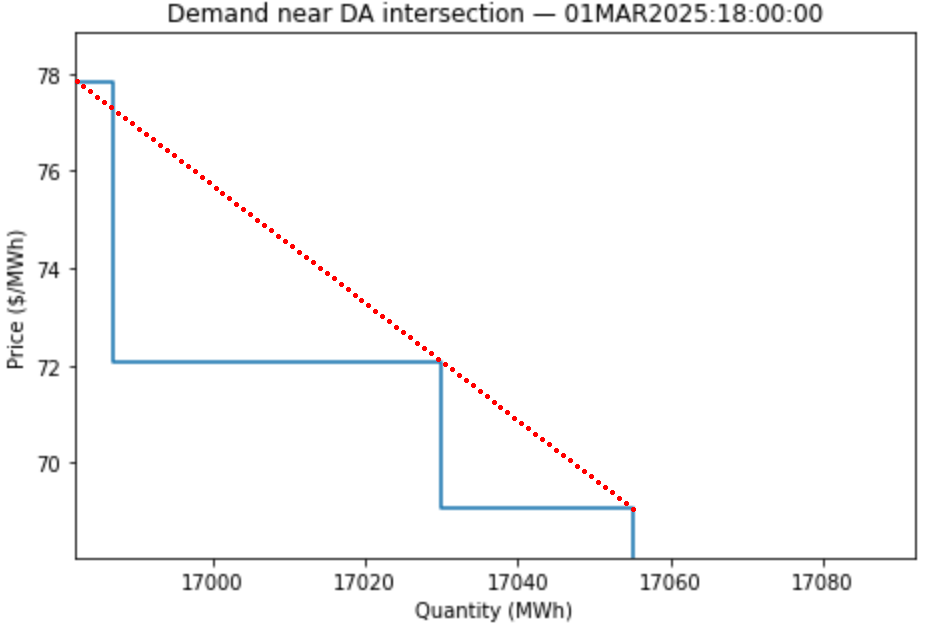}
    \caption{03-01-2025}
    \label{fig:zoom-0501}
  \end{subfigure}\hfill
  \begin{subfigure}[t]{0.32\textwidth}
    \centering
    \includegraphics[width=\linewidth]{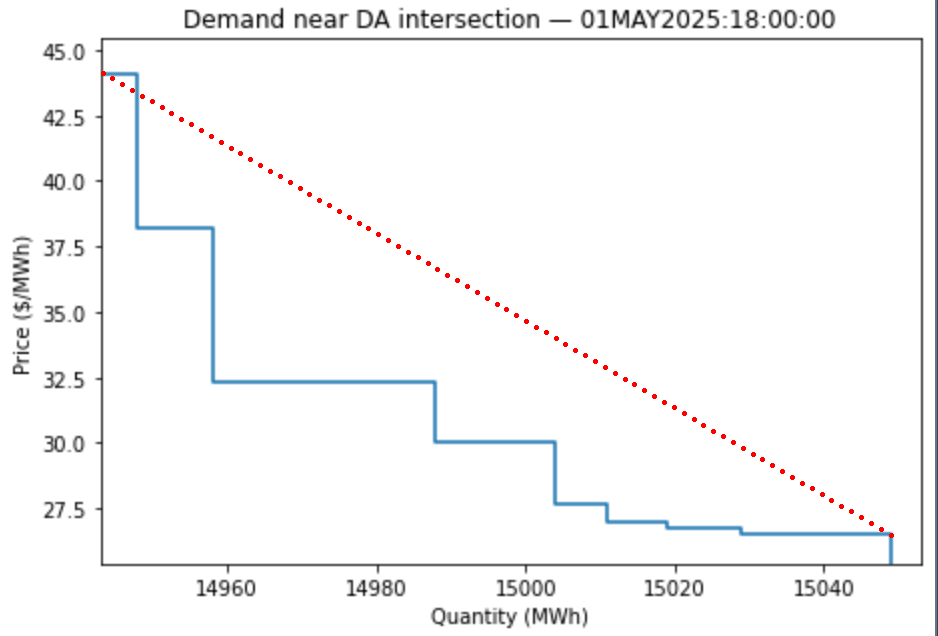}
    \caption{05-01-2025}
    \label{fig:zoom-0301}
  \end{subfigure}\hfill
  \begin{subfigure}[t]{0.32\textwidth}
    \centering
    \includegraphics[width=\linewidth]{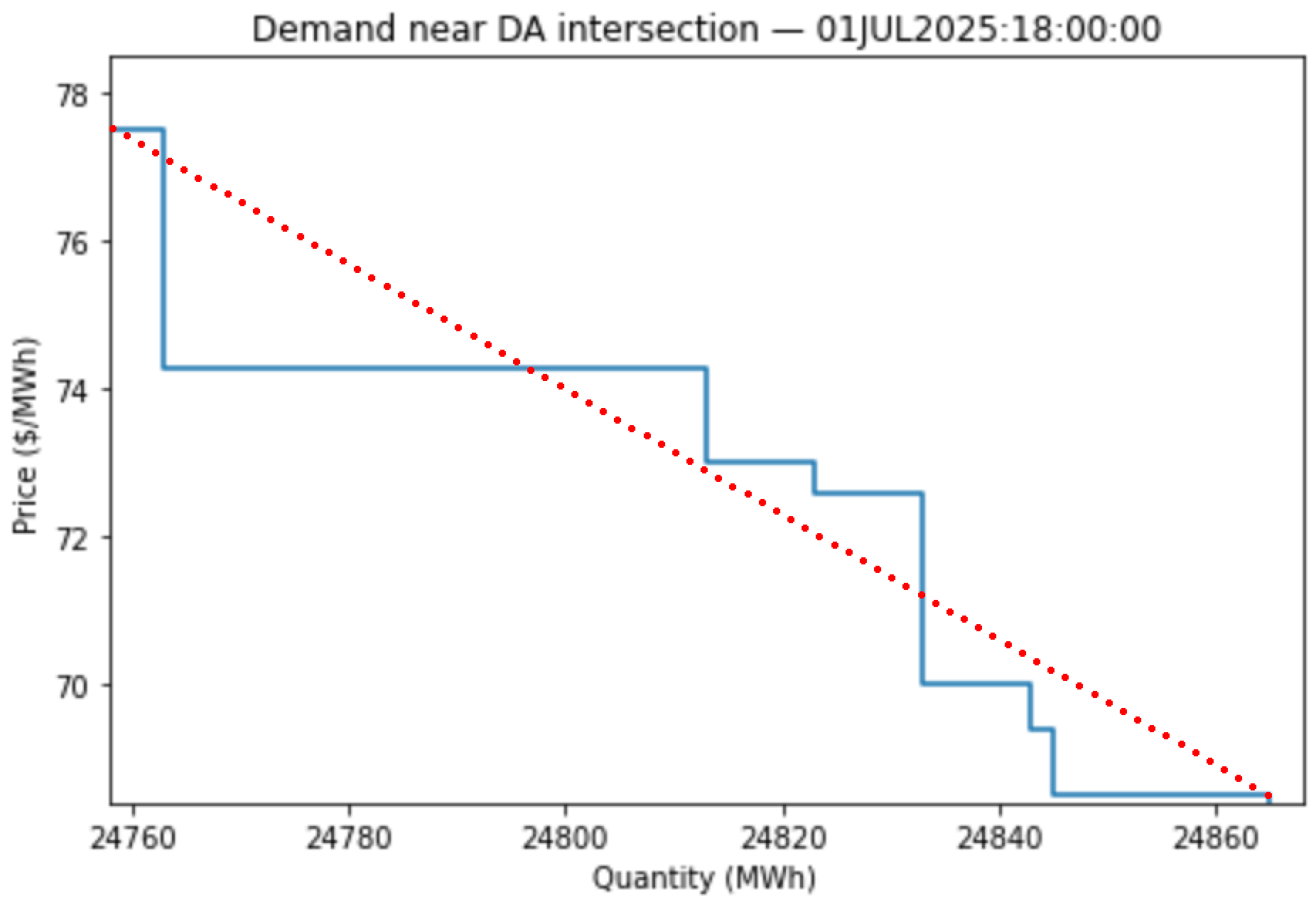}
    \caption{07-01-2025}
    \label{fig:zoom-0701}
  \end{subfigure}
  \caption{{Demand stack and linear approximation near the DA price-setting intersection point at three different hours}.}
  \label{fig:zoomed-demand-row}
\end{figure}

\revision{The previous derivation shows that the linear price impact specification arises naturally as a first-order expansion of the aggregate supply (resp. demand) curve around the day-ahead market-clearing point. Such an approximation is valid precisely because virtual traders submit small quantities relative to total cleared volume, so that the perturbed clearing price $p^+$ remains close to $p^\star$. This approximation is further supported empirically: Figures~\ref{fig:zoomed-supply-row} and~\ref{fig:zoomed-demand-row} show that the bid stack is well-approximated by a linear slope in the neighborhood of the equilibrium across a range of representative hours, seasons, and load regimes, and Table~\ref{tab:impact-24jun-2025} confirms that the realized price perturbations on the most profitable day of the sample are fully consistent with the calibrated linear coefficients.}

\revision{In practice, $(Q^{\rm S})'(p^\star)$ is not directly observable, as the bid stack is a discrete collection of bids rather than a smooth analytical function. We therefore estimate this derivative numerically using one-sided finite differences. This approach follows the bid-stack-based price formation framework of \cite{howison2009stochastic}, which models electricity prices as arising from local properties of the aggregate supply and demand curves.}

\revision{We now apply this approach to the NYISO market to estimate $k_{\rm E}^{+}$. In practice, the energy-impact coefficient is time-dependent: the slope of the supply curve varies systematically with the season and the time of day, reflecting predictable changes in the bid-stack geometry driven by demand patterns and generation mix. To account for this, we stratify the estimation by season (Winter, Summer, Shoulder) and load regime (Peak and Off-Peak hours). Within each stratum, we take $\Delta q = 1000$~MWh and average the resulting finite-difference price responses across the Top-10 spike hours. By focusing on the hours with the largest price spikes, where supply curves tend to be steepest, this procedure deliberately yields a conservative estimate of $k_{\rm E}^{+}$: the resulting coefficient overestimates the price impact in typical hours, which in turn leads to more cautious position sizing and a lower risk of underestimating market impact costs. The resulting average buy-side impacts are reported in Table~\ref{tab:mean-impact-by-bucket-longil}.}

\begin{table}[H]
\centering
\caption{Average linear price impact $k_{\rm E}^+( \$/1000$MWh) induced by a $+1000$ MWh demand shock, by season and load band: train (2015--2019) vs.\ test (2022--2025).
}
\label{tab:mean-impact-by-bucket-longil}
\begin{tabular}{l l S[table-format=3.3] S[table-format=3.3]}
\toprule
Season & Band & {Top-10 spikes (2015--2019)} & {Top-10 spikes (2022--2025)} \\
\midrule
Winter   & Off-Peak & 17.030 & 10.561 \\
Winter   & Peak     & 23.210 & 26.270 \\
Summer   & Off-Peak &  8.050 & 66.839 \\
Summer   & Peak     & 34.640 & 46.477 \\
Shoulder & Off-Peak &  1.740 &  2.342 \\
Shoulder & Peak     & 10.480 & 22.226 \\
\bottomrule
\end{tabular}
\end{table}

Similarly, a net short position of size $\Delta q<0$ corresponds to a shift
along the inverse demand curve. In this case, the perturbed clearing price $p^-$
satisfies
\begin{equation}
\label{taylor2}
p^- - p^\star \;\approx\; \frac{\Delta q}{(Q^{\rm D})'(p^\star)} \;<\; 0.
\end{equation}
The sell-side impact is thus governed by the local slope of the aggregate demand
curve. We estimate $k_{\rm E}^{-}$ analogously using one-sided finite differences with
$\Delta q=1000$~MWh and the same sampling scheme.
Corresponding sell-side impacts are reported in
Table~\ref{tab:deltaP_sell1000_combined}.

The parameters $(k_{\rm E}^{+},k_{\rm E}^{-})$ should therefore be interpreted as local average slopes of the inverse residual supply and demand curves at the day-ahead clearing point. In general, they need not coincide: supply and demand can exhibit markedly different slopes near equilibrium, particularly during stressed system conditions. This naturally leads to an asymmetric price response to buy- versus sell-side shocks. For illustration, in the calibration sample (2015--2019) the largest buy-side impact occurs during {Summer Peak} hours, where a $+1000$~MWh demand shock raises prices by $34.64$~\$/MWh on average, while on the sell side, the largest impact is observed during {Winter Peak} hours, where a $-1000$~MWh supply shock lowers prices by $114.89$~\$/MWh.

\begin{table}[H]
\centering
\caption{Average linear price impact $k_{\rm E}^-( \$/1000$MWh) induced by a $-1000$ MWh supply shock, by season and load band: train (2015--2019) vs.\ test (2022--2025).
}
\label{tab:deltaP_sell1000_combined}
\begin{tabular}{l l S[table-format=4.2] S[table-format=4.2]}
\toprule
Season & Band & {Top-10 spikes (2015--2019)} & {Top-10 spikes (2022--2025)} \\
\midrule
Winter   & Off-Peak & -92.15  & -117.95 \\
Winter   & Peak     & -114.89 & -131.33 \\
Summer   & Off-Peak & -19.73  &  -24.40 \\
Summer   & Peak     & -45.82  &  -59.11 \\
Shoulder & Off-Peak & -17.96  &  -22.78 \\
Shoulder & Peak     & -28.37  &  -32.44 \\
\bottomrule
\end{tabular}
\end{table}

\revision{Tables~\ref{tab:mean-impact-by-bucket-longil} and \ref{tab:deltaP_sell1000_combined} also reveal substantial regime-dependent variation in the energy-impact coefficients: buy-side impacts are systematically larger during Summer Peak hours, reflecting 
heightened system stress and steeper supply curves, while sell-side impacts 
are most pronounced during Winter Peak hours, consistent with tighter demand 
conditions and less elastic demand curves in cold weather. This seasonal and 
load-regime variation in price sensitivity is an economically meaningful 
feature of market microstructure that extends beyond its role in portfolio 
optimization.}

Finally, Figures~\ref{fig:zoomed-supply-row} and~\ref{fig:zoomed-demand-row} provide a visual illustration of the local linearity underlying this calibration. For a selection of representative hours, we plot the supply and demand curves in a neighborhood of the Day-Ahead Market intersection and compare the market-clearing price before and after an exogenous quantity shock. The dotted line in each panel corresponds to the local linear approximation used in the impact model, while the displacement between the pre- and post-shock clearing prices reflects the realized price response to the injected demand or supply. These examples illustrate how, at the relevant operating point, the bid stack is well-approximated by a linear slope, justifying the use of a linear impact specification and motivating the estimation of seasonal and load-regime–specific impact coefficients.

In all subsequent optimization experiments, the system-wide impact parameters $(k_{\rm E}^{+}, k_{\rm E}^{-})$ are estimated exclusively on the calibration sample (2015--2019) and treated as fixed inputs when constructing optimal portfolios and computing realized P\&L over the out-of-sample test period (2022--2025). This procedure ensures that portfolio decisions rely solely on historically available information and that all reported profits are free of look-ahead bias.

\subsection{Estimating local impact coefficients}\label{sec:estimating-kz}

\revision{The zone-specific impact coefficients $k_z$ capture the marginal response of the loss and congestion components of the locational marginal price (LMP) to incremental changes in zonal demand. Congestion and losses reflect the marginal cost of delivering electricity across a constrained network, so the coefficients $k_z$ can be interpreted as local approximations of these marginal costs. They therefore provide a reduced-form measure of how localized demand shocks propagate through the transmission network and affect price formation, with implications beyond virtual trading.}

\revision{To estimate these coefficients, we regress the DA loss-minus-congestion component on the corresponding zonal forecast load separately for each zone, season, and Peak/Off-Peak bucket.} As a reference, we then focus on the Long Island (LONGIL) zone, which is both a large demand center and an import-constrained load pocket in NYISO. Its loss and congestion components reflect flows across multiple upstream interfaces and exhibit substantial non-local transmission stress, making Long Island a natural baseline for calibrating marginal price impact. 

\revision{The estimated slopes for NYISO, reported in Table \ref{tab:impact-loss-minus-cong-new},} measure the average price impact (in \$/MWh) of a $+1000$~MWh increase in forecast load within that zone. In Long Island, the estimated impact ranges from $4.95$ to $7.82$ in Shoulder, from $5.06$ to $17.73$ in Summer, and from $43.30$ to $43.63$ in Winter, depending on the Peak/Off-Peak bucket. Table~\ref{tab:avg-load-2015-2021} reports the corresponding average forecast load by zone and season over 2015--2021. Zones with higher typical loads (e.g., NYC and Long Island) tend to exhibit smaller marginal price impacts per~MW, whereas smaller or more transmission-constrained zones (e.g., Millwood and Dunwoodie) show substantially higher sensitivities. These patterns reflect the dilution of localized load shocks in large demand centers and their amplification in smaller or more constrained zones.

\revision{Comparing the regression estimates in Table~\ref{tab:impact-loss-minus-cong-new} with the average forecast loads reported in Table~\ref{tab:avg-load-2015-2021}, we observe that zones with higher average loads tend to exhibit smaller marginal price impacts. This pattern is consistent with the findings of \cite{NYISO_LCR_2024_2025, NYISO_Congestion_Component_2025}, and suggests modeling the local impact coefficients $k_z$ as inversely proportional to average zonal load:}
\[
k_z = k_{\mathrm{LONGIL}} \frac{L_{\mathrm{LONGIL}}}{L_z},
\qquad k_{\mathrm{LONGIL}} = 0.050\$/(\text{MWh})^2,
\]
\revision{where $L_z$ denotes the historical mean actual load in zone $z$ over 2015--2021. The scale of the coefficients is anchored to the Long Island Summer--Peak regression estimates (shown in Figure~\ref{fig:random-100}), while preserving the empirical ranking of zones by size and sensitivity.} This yields the following estimates:
\begin{align*}
k_{\text{NYC}}    & = 0.020 \quad &
k_{\text{LONGIL}} & = 0.050 \quad &
k_{\text{WEST}}   & = 0.067 \quad &
k_{\text{CENTRL}} & = 0.065, \\
k_{\text{CAPITL}} & = 0.085 \quad &
k_{\text{NORTH}}  & = 0.210 \quad &
k_{\text{DUNWOD}} & = 0.169 \quad &
k_{\text{MILLWD}} & = 0.357, \\
k_{\text{HUDVL}}  & = 0.105 \quad &
k_{\text{MHKVL}}  & = 0.129 \quad &
k_{\text{GENESE}} & = 0.103. \quad 
\end{align*}
Zones with larger average loads (e.g., NYC, LONGIL, CENTRL) therefore accommodate larger virtual positions with lower price impact, while smaller or more constrained zones (e.g., MILLWD and NORTH) exhibit higher sensitivity. 

\revision{This specification should be interpreted as a reduced-form approximation of local price sensitivity, which remains parsimonious while preserving the empirical ranking of zones by size and impact. Since optimal allocations are driven by large zones such as Long Island and NYC, the results are not highly sensitive to moderate variations in smaller zones, as confirmed by the robustness checks in Section~\ref{sec:robustness}. The implied scale is also consistent with empirical limits, for example around 50~\$/1000~MWh in Long Island and around 15--20~\$/1000~MWh in NYC. Overall, this calibration highlights that zonal price impact is inherently heterogeneous and reflects underlying differences in congestion topology and network constraints across the system.}

Finally, Table~\ref{tab:avg_corr_load_cong_loss} reports average correlations between forecast load and the day-ahead loss and congestion components across zones. Both components are consistently positively correlated with forecast load across all seasons and time buckets, supporting the modeling assumption that higher forecasted load increases losses and congestion in expectation. 

\section{Performance in NYISO}\label{sec:performance}

\revision{This section evaluates the performance of the 
optimal trading strategy developed in Section~\ref{sec:scaling-impact}, 
and contrasts it with the benchmark unit-size strategies 
of Section~\ref{benchstrats}. While Section~\ref{benchstrats} 
assessed the predictive performance of the spike-forecasting 
model under fixed unit-size positions and no price impact, 
the results reported here reflect the realized interaction 
between forecasting accuracy, market impact, and cross-zonal 
portfolio allocation. All predictive models, impact parameters, 
and decision thresholds are fixed based on the calibration 
and validation samples, and performance is assessed 
out-of-sample on NYISO data from 2022--2025.}

\subsection{Main results}

\revision{We begin by assessing the performance of the predictive model on the validation period (2020--2021), which serves two purposes: it allows us to select the optimal probability thresholds $\tau_z$ for each zone, and it provides a basis for filtering out zones where the model fails to generate economically meaningful signals.}

\revision{Table~\ref{tab:validation_summary} reports, for each zone, the number of trades and average P\&L per trade on the validation period. The results reveal substantial heterogeneity across zones: while some zones generate consistently positive average P\&L, others exhibit negative or near-zero performance, suggesting that the predictive model lacks sufficient discriminatory power in those zones. To avoid carrying forward signals that are not economically validated, we exclude any zone for which the average out-of-sample P\&L on the validation period is negative. This conservative filtering criterion ensures that only zones where the model has demonstrated predictive value are included in the test period, reducing the risk of overfitting to noise. As reported in Table~\ref{tab:validation_summary}, this leads us to exclude the North zone for INC predictions and the Long Island zone for DEC predictions.}

\begin{table}[H]
\centering
\small
\caption{Validation period (2020--2021): trade counts, average P\&L (USD/MWh), and eligibility per zone.}
\label{tab:validation_summary}
\begin{tabular}{lrrrrrr}
\toprule
\textbf{Zone} & \textbf{DEC Trades} & \textbf{DEC Avg Win} & \textbf{INC} & \textbf{INC Avg Win}   \\
\midrule
NYC    & 122 & 12.35     & 326 & 3.46    \\
LONGIL & 316 & 38.12     & 1705 & -1.07  \\
WEST   & 152 & 5.21      & 307 & 9.76     \\
CENTRL & 36  & 2.26     & 61  & 17.41    \\
CAPITL & 75  & 0.63      & 206 & 11.00    \\
NORTH  & 41  & -0.13    & 25  & 6.48     \\
DUNWOD & 49  & 7.72      & 189 & 3.03    \\
MILLWD & 56  & 18.81     & 167 & 0.82   \\
HUDVL  & 47  & 3.30      & 159 & 9.92    \\
MHKVL  & 33  & 2.65     & 52  & 16.47   \\
GENESE & 40  & 12.25     & 51  & 13.75  \\
\bottomrule
\end{tabular}
\end{table}

\revision{Having identified the zones with validated predictive signals, we now apply the optimal trading strategy of Section~\ref{sec:scaling-impact} to determine position sizes. A key feature of this strategy is that trades are determined jointly across zones, rather than independently for each zone. More precisely,} while the predictive model assigns to every zone--hour a directional signal (DEC for negative predicted DART and INC for positive predicted DART), the executed trades are determined jointly across zones. In particular, it may allocate a position whose sign differs from the local predicted direction. For example, it can place a DEC in one zone while taking an INC in another, even when both zones individually exhibit positive expected DART spreads. This behavior is driven by the system-wide impact penalty $k_{\rm E}$: by taking offsetting positions, the strategy reduces aggregate exposure and associated costs, allowing risk to be concentrated in zones where marginal price impact is lowest.

\begin{figure}[ht]
  \centering
  \includegraphics[width=\linewidth]{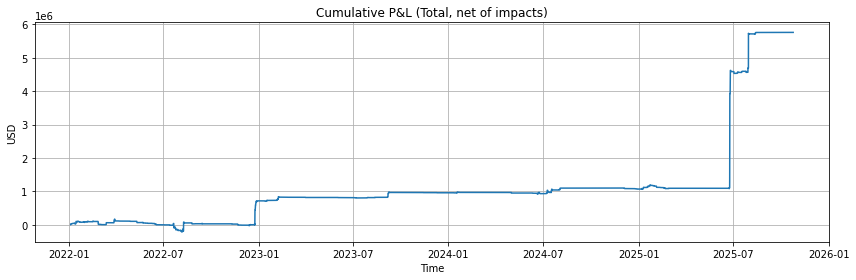}
  \vspace{0.6em}

  \begin{subfigure}{0.49\linewidth}
    \centering
    \includegraphics[width=\linewidth,height=0.13\textheight]{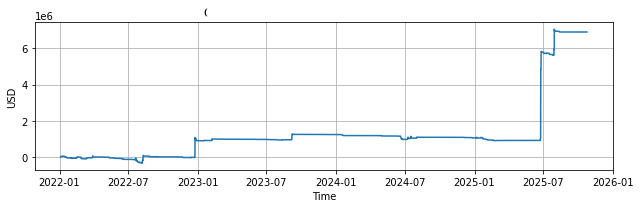}
    \caption{DEC — execution view}
  \end{subfigure}\hfill
  \begin{subfigure}{0.49\linewidth}
    \centering
    \includegraphics[width=\linewidth,height=0.13\textheight]{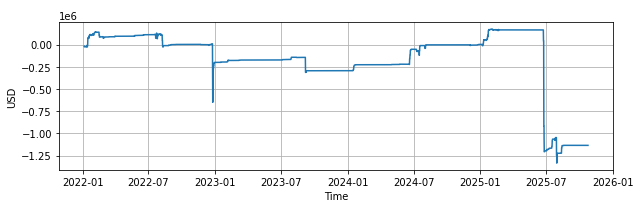}
    \caption{INC — execution view}
  \end{subfigure}

  \begin{subfigure}{0.49\linewidth}
    \centering
    \includegraphics[width=\linewidth,height=0.13\textheight]{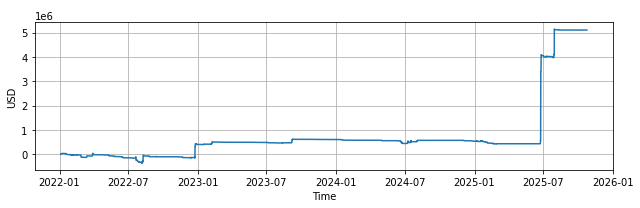}
    \caption{DEC — prediction view}
  \end{subfigure}\hfill
  \begin{subfigure}{0.49\linewidth}
    \centering
    \includegraphics[width=\linewidth,height=0.13\textheight]{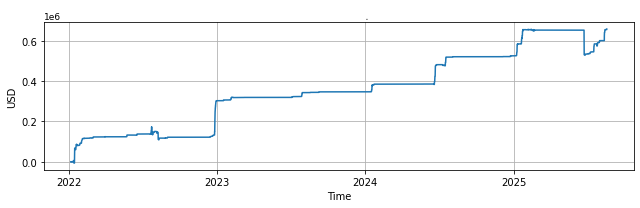}
    \caption{INC — prediction view}
  \end{subfigure}

  \caption{Cumulative P\&L: total (top) and by side and attribution view (bottom). All values are in USD.}
  \label{fig:pnl_panels}
\end{figure}

As a result, a zone may exhibit negative realized P\&L despite a correct directional prediction. To disentangle model accuracy from portfolio allocation effects, we therefore report the results over the 2022--2025 test period in Figure~\ref{fig:pnl_panels} in two complementary views: the {prediction view} evaluates performance based solely on the model’s directional signals, while the {execution view} reflects realized P\&L after optimization and cross-zone balancing. \revision{The top panel shows the total cumulative P\&L net of price impact, which reaches approximately \$5.76 million by the end of 2025, with the majority of gains concentrated in the second half of 2025. The bottom panels decompose the P\&L by side and attribution view: the DEC execution view (panel a) drives most of the total P\&L, while the INC execution view (panel b) is slightly negative, reflecting the contra-directional positions taken by the optimizer to reduce system-wide impact. The prediction views (panels c and d) are both positive, confirming that the model's directional signals are economically meaningful, and that the negative INC execution P\&L is an artifact of portfolio optimization rather than poor predictive performance.}

\revision{A striking feature of the cumulative P\&L is its concentration in a single episode: the sharp jump visible in the top panel of Figure~\ref{fig:pnl_panels} in late June 2025 corresponds to a heat-wave event on 24~June~2025. On that day alone, the strategy generated a substantial fraction of its total out-of-sample P\&L, driven by unusually large DART dislocations in constrained downstate zones. This concentration is consistent with the interpretation of the strategy's profits as compensation for bearing tail risk: large gains arise precisely when system stress is highest and the model's predictions are most valuable.}

\revision{This episode also provides a natural opportunity to validate the calibration of the energy-impact coefficients $k_{\rm E}$. On that day, the strategy repeatedly submits large net buy (DEC) positions during Peak hours, making it an ideal setting to compare predicted and realized price impacts.} Using the NYISO bid stack for these same hours, we recompute the day-ahead clearing price after shifting the residual demand curve upward by the executed quantity $q_t$ MWh, measuring the realized one-sided price impact
\[
\Delta P_t = P^{\mathrm{DA}}_t(q_t) - P^{\mathrm{DA}}_t,
\]
and the corresponding empirical slope
\[
\frac{\Delta P_t}{q_t} \times 1000 \qquad 
(\text{\$/MWh per 1000 MWh}).
\]
The results in Table~\ref{tab:impact-24jun-2025} show that the realized slopes lie between roughly $8$ and $41$\$/MWh per 1000~MWh, fully consistent with the calibrated Summer-Peak coefficients obtained in Section~\ref{sec:modelling_ke}, namely $k_{\rm E}^{+} \in [34.64, 46.48]$ \$/1000~MWh. The empirical price impact observed during the largest trading day thus strongly supports the validity of our linear approximation and our chosen parameter values.

Table~\ref{tab:test_zone_execution} reports, for each zone, the number of active trading hours, the average absolute position size, and the resulting P\&L. The largest virtual positions are taken in Long Island, which is consistent with its market characteristics: the zone combines frequent and sizable DART spikes with high average load, implying a relatively low marginal price impact per traded MWh. Finally, Table~\ref{tab:zone_exec_agg} reports zone-level performance, while Table~\ref{tab:yearly_totals_small} aggregates results across all zones for each test year. 

\begin{table}[H]
\centering
\small
\caption{Per-zone attribution on the TEST period (2022--2025), execution view with dynamic $q^\star$ and price impacts.}
\label{tab:test_zone_execution}
\begin{tabular}{lrrrrr}
\toprule
\textbf{Zone} & \textbf{Hours Active} & \textbf{Avg $|q|$ (MW)} & \textbf{P\&L (USD)}  \\
\midrule
LONGIL & 571  & 118.70 & 6{,}004{,}480 \\
NYC    & 1009 & 17.71  &   722{,}617  \\
MILLWD & 775  & 2.67   &   119{,}145 \\
GENESE & 397  & 8.91   &   118{,}976   \\
WEST   & 1140 & 12.34  &    26{,}839   \\
NORTH  & 160  & 3.32   &   -10{,}806   \\
MHKVL  & 426  & 12.64  &  -101{,}829  \\
DUNWOD & 803  & 3.15   &  -169{,}380  \\
CENTRL & 440  & 27.54  &  -209{,}366  \\
CAPITL & 1026 & 14.34  &  -355{,}918  \\
HUDVL  & 733  & 9.10   &  -384{,}604   \\
\midrule
\textbf{Total / Stats:} & & & \textbf{5{,}760{,}154} & & \\
\bottomrule
\end{tabular}
\end{table}

\subsection{Forecasting diagnostics}
\label{sec:spike_forecasting}
On the 2022--2025 test period, the joint model exhibits a highly selective forecasting behavior, characterized by high precision but low recall, \revision{as reported in Tables~\ref{tab:test_zone_metrics_inc} and~\ref{tab:test_zone_metrics_dec}}. Average precision is approximately $0.30$ for DEC trades and $0.77$ for INC trades, indicating relatively few false positives, while recall remains around $0.04$--$0.05$, meaning that only a small fraction of all realized spike events is acted upon. This pattern reflects a profit-oriented design: the strategy trades only when the model assigns high confidence to extreme DART deviations, favoring signal quality over coverage.

To assess whether the model is selecting {economically relevant} spikes rather than arbitrary subsets of hours, Figures~\ref{fig:hour_histograms} and~\ref{fig:month_histograms} compare the empirical distributions of realized DART spikes with those selected by the model. In each case, we compare both against the full sample of hours and against the largest spike quantiles. Across both hour-of-day and month-of-year dimensions, the model-selected distributions align much more closely with the largest observed spikes than with the unconditional sample.

This effect is quantified using the Jensen--Shannon (JS) divergence, a symmetric and bounded measure of dissimilarity between two probability distributions. Given distributions $P$ and $Q$ on a common support, define their mixture $M=\tfrac{1}{2}(P+Q)$. The JS divergence is
\[
\mathrm{JS}(P\|Q)
=
\frac{1}{2}\,\mathrm{KL}\!\left(P\middle\|M\right)
+
\frac{1}{2}\,\mathrm{KL}\!\left(Q\middle\|M\right),
\]
where $\mathrm{KL}(P\|Q)=\sum_x P(x)\log\!\big(\tfrac{P(x)}{Q(x)}\big)$ denotes the Kullback--Leibler divergence. We note that $\mathrm{JS}(P\|Q)\in[0,\log 2]$, with smaller values indicating more similar distributions \cite{Lin1991}. In particular,
\[
\textbf{DEC:}\quad \mathrm{JS}_{\text{top }20\%}=0.039
\quad \text{vs.} \quad
\mathrm{JS}_{\text{all}}=0.094,
\]
\[
\textbf{INC:}\quad \mathrm{JS}_{\text{top }5\%}=0.061
\quad \text{vs.} \quad
\mathrm{JS}_{\text{all}}=0.116.
\]
The substantially smaller divergences for the top spike quantiles show that predicted trading hours are statistically closer to the most extreme realized price deviations than to typical hours. In other words, although the model captures only a small fraction of all spikes, it disproportionately targets events that resemble the largest historical DART dislocations, consistent with the subsequent profitability of the trading strategy.
 
\subsection{Strategy variants}

In the baseline strategy, the portfolio optimizer is free to choose the sign of the zonal position \(q_{t,z}\) so long as the resulting expected revenue (after impact) is positive. As a consequence, a zone can end up with an executed INC position even in hours where the spike–forecasting model indicates a negative DART (and hence a DEC position), or vice versa, because the optimizer may use offsetting positions across zones to reduce system-wide impact costs. 

To enforce a tighter link between forecasts and execution, we introduce a {side-clipping} rule : at each zone–hour we retain only the signal consistent with the predicted DART sign (DEC if the model predicts negative DART, INC if it predicts positive DART), and set all conflicting signals to zero before testing the strategy. This mechanism forces each traded hour to take positions only in the direction implied by the model’s sign prediction, while still allowing the optimizer to choose the trade size.

Figure~\ref{fig:clip_results} shows that this clipped strategy performs better on the 2022--2025 test set. The total P\&L increases relative to the unconstrained optimizer, and both the INC and DEC components become smoother and less noisy. Table~\ref{tab:perzone_exec_plain} reports the per-zone attribution of the clipped strategy. Restricting trades to the model-consistent direction reduces the number of active hours in some zones, but the remaining positions tend to have higher average profitability. Long Island remains the dominant contributor, followed by New York City and Capital, reflecting the strength of DEC trades in these load pockets.

Overall, the clipped strategy preserves most of the performance of the full optimizer while improving interpretability and robustness. By aligning trades strictly with the predicted sign of DART, it reduces contradictory positions and produces a cleaner mapping between forecasts and executed trades.

Building on this clipped strategy, we then introduce a further refinement by restricting attention to zone--season--band buckets that exhibit statistically significant performance in the validation period (2020--2021). Specifically, we split the analysis by seasonality (Winter/Summer/Shoulder) and Peak/Off-Peak hours, and retain only those buckets whose mean P\&L has a $t$-statistic exceeding~$2$ and the number of trades is at least $50$, corresponding approximately to 95\% confidence against a zero-mean null.

Under this restriction, the universe of traded zones shrinks substantially, as Tables \ref{tab:inc_valid_zone_season_band} and \ref{tab:dec_valid_zone_season_band} suggest. On the DEC side, only Long Island satisfies the significance criterion, whereas on the INC side the retained zones are Capital, Central, Long Island, New York City, and West. All other zone--band combinations are excluded from trading in the test period.

Figure~\ref{fig:pnl-panels} reports cumulative P\&L over 2022--2025 for this restricted strategy. As expected, total profits are slightly lower than in the fully pooled specification, reflecting the reduced number of traded positions. Nevertheless, performance remains strong, with the majority of gains concentrated during the large summer 2025 spike episodes.

To assess predictive accuracy, Table~\ref{tab:pred-quality} summarizes the fraction of trades that coincide with realized spikes and the frequency with which the sign of the DART spread is correctly predicted. On the DEC side, approximately 27\% of trades occur during realized positive spikes, and the model predicts the correct sign in 41\% of hours. By contrast, INC predictions are substantially more reliable: roughly 76\% of DEC trades coincide with negative spikes, and nearly 80\% correctly predict the spread sign.

\revision{Overall, this analysis highlights an asymmetry between INC and DEC signals, consistent with the evidence presented in Section~\ref{sec:spike_forecasting}. While DEC opportunities are relatively sparse and more difficult to identify, INC signals exhibit stronger persistence and higher predictive accuracy. This pattern is consistent with underlying market dynamics, where negative DART realizations are more systematically associated with predictable congestion and demand conditions.}

{\color{black}
\subsection{Sensitivity and robustness analysis}\label{sec:robustness}

We finally assess the robustness of the main results to alternative modeling choices, considering in turn the calibration of the local impact coefficients, the specification of spike thresholds, and the choice of classification method.

\paragraph{Robustness to alternative impact calibration.} We also consider an alternative specification in which the local impact coefficients $k_z$ are estimated directly from data via linear regression of Table~\ref{tab:impact-loss-minus-cong-new}, rather than being scaled relative to the Long Island zone. The resulting coefficients are of comparable magnitude across zones and remain within the empirical range observed in Section~\ref{sec:modelling_ke}. Re-estimating the strategy under this specification yields very similar out-of-sample performance, as Figure~\ref{fig:pnl_new_k} suggests. In particular, the aggregate P\&L remains stable, with an increase of approximately one million USD. This is reasonable, since the impact coefficients are lower in most of the zones, so the strategy can be scaled to a greater degree. Risk-adjusted metrics are largely unchanged: the annualized Sharpe ratio is approximately 0.85, and the maximum drawdown is about -450{,}000 USD. This suggests that the performance of the strategy is robust to the precise calibration of impact coefficients and is not driven by a particular normalization choice.

\paragraph{Sensitivity check.} As a sensitivity analysis, we varied the spike thresholds used to define INC and DEC events, taking $\gamma_{\text{neg}} \in \{5,10,15,30,45,60\}$ and $\gamma_{\text{pos}} \in \{2,5,8,10,15,25\}$, and recomputed the resulting total out-of-sample P\&L in each case. The corresponding results are reported in Table~\ref{tab:gamma_sensitivity} below. The results indicate that total out-of-sample P\&L is maximized around the benchmark threshold pair \((\gamma^+,\gamma^-)=(5,30)\), and tends to decline as one moves sufficiently far away from this region in either direction.

\begin{table}[H]
\centering
\small
\caption{Sensitivity of total out-of-sample P\&L (USD) to the spike-threshold specification in NYISO. Rows correspond to $\gamma^-_{\text{neg}}$ and columns to $\gamma^+_{\text{pos}}$.}
\label{tab:gamma_sensitivity}
\begin{tabular}{c|rrrrrr}
\toprule
$\gamma^-_{\text{neg}} \backslash \gamma^+_{\text{pos}}$ & 2 & 5 & 8 & 10 & 15 & 25 \\
\midrule
5  & 4{,}239{,}207 & 4{,}426{,}372 & 4{,}553{,}582 & 3{,}743{,}014 & 3{,}437{,}603 & 3{,}060{,}061 \\
10 & 4{,}403{,}255 & 4{,}525{,}201 & 4{,}491{,}844 & 3{,}779{,}073 & 3{,}529{,}344 & 3{,}089{,}127 \\
15 & 5{,}306{,}516 & 5{,}436{,}941 & 5{,}420{,}987 & 4{,}609{,}272 & 4{,}478{,}300 & 3{,}917{,}606 \\
30 & 5{,}598{,}509 & 5{,}760{,}153 & 5{,}514{,}463 & 4{,}780{,}307 & 4{,}222{,}773 & 4{,}002{,}362 \\
45 & 4{,}843{,}657 & 4{,}926{,}858 & 4{,}857{,}169 & 4{,}136{,}319 & 4{,}057{,}745 & 3{,}439{,}808 \\
60 & 4{,}667{,}539 & 4{,}760{,}067 & 4{,}677{,}236 & 3{,}962{,}591 & 3{,}937{,}211 & 3{,}263{,}881 \\
\bottomrule
\end{tabular}
\end{table}

\paragraph{Robustness to classification method.} As an additional robustness check, we experimented with SMOTE-based resampling to address the rarity of spike events in the training data. While SMOTE generates synthetic samples by interpolating between observed spike realizations, such configurations may not be economically meaningful, introducing noise rather than genuine predictive signal. Empirically, the resampled specification generated more trading signals but with lower average profitability per trade---for example, in the NYC zone, total P\&L decreases from approximately \$6,000,000 to around \$3,000,000 when SMOTE is applied. We therefore retain the baseline logistic specification without resampling.}

\section{Conclusion}
\label{sec:conclusion}

\paragraph{Main contributions.} \revision{This paper develops a unified framework for forecasting and optimally trading day-ahead versus real-time (DART) price spreads in U.S. wholesale electricity markets. The central insight is that forecasting and trading must be treated as a joint problem: predictive signals alone are insufficient unless paired with a principled model of price impact and a rigorous position-sizing rule. Building on the framework of \cite{galarneau2022foreseeing}, we extend spike prediction from a single zone to a multi-zone setting, treat both positive and negative DART spikes within a unified model, and go beyond unit-size trading by developing a structural price-impact model that yields closed-form expressions for the optimal vector of zonal positions.}

Working with NYISO, ISO--NE, and ERCOT, we first construct leakage-free feature sets that respect each ISO's day-ahead bid deadline and estimate zone-specific logistic regressions for both positive and negative DART spikes. The resulting models are deliberately selective: they trade only when assigned high spike probabilities, thereby prioritizing economic relevance over statistical coverage.Then, using historical day-ahead bid stacks, we estimate system-wide energy impact coefficients and zone-specific congestion sensitivities, yielding a linear-quadratic impact model for virtual load. Closed-form expressions for the optimal zonal quantities show that portfolio-level decisions are shaped not only by local expected revenues, but also by cross-zone interactions through the common energy component. In particular, it can be optimal to take offsetting positions across zones in order to concentrate risk where marginal price impact is lowest. Backtests that ignore this feedback either overstate achievable profits or implicitly assume unrealistically small position sizes. \revision{This highlights that trading profits are fundamentally constrained by market clearing mechanisms: any profitable strategy must account for the endogenous feedback between submitted quantities and market-clearing prices.}

\paragraph{Empirical findings.} Our empirical results first show that spike predictability is highly heterogeneous across markets and zones. In ISO--NE and ERCOT, DART spreads are almost perfectly synchronized across load zones, so a single representative node captures nearly all useful variation. By contrast, NYISO exhibits much weaker and more dispersed cross-zonal correlations, driven by localized congestion, \revision{transmission constraints, and heterogeneous marginal costs across the network}. In this environment, multi-zone modelling is essential: different zones deliver structurally different DART distributions and support distinct INC/DEC opportunities. At the zone level, Long Island emerges as the most profitable load pocket, combining frequent extreme spreads with high average load, while several upstate zones offer weaker but still statistically significant signals.

From a trading perspective, the most stable opportunities arise on the INC side. Positive DART spikes (which generate INC profits) occur more frequently and with greater statistical regularity, and the forecasting model identifies these events with higher precision. In contrast, DEC opportunities, driven by negative DART spikes, are substantially more lucrative when they occur, but they are rarer, more volatile, and more sensitive to threshold selection and market-impact assumptions. As a result, INC strategies deliver smoother and more persistent returns, whereas DEC strategies contribute occasional but very large profit bursts. Restricting trades to directions consistent with the model's sign prediction and to statistically significant zone-season-band buckets preserves most of the economic value while improving robustness and interpretability.

\paragraph{Economic interpretation.} \revision{The profitability of the strategy raises a natural question: do these profits reflect risk premia, persistent market inefficiencies, or compensation for liquidity provision? Empirically, DART spreads reflect market adjustments to congestion, forecast errors, and system imbalances, and can be interpreted as a market-implied forecast error up to an embedded risk premium in day-ahead prices \cite{longstaff2004electricity}. This suggests that virtual traders who correctly anticipate extreme spreads are intermediating these imbalances and bearing short-term pricing risk. Under this interpretation, several features of our results are consistent with a risk premium view. First, the asymmetry between INC and DEC opportunities suggests that downward price corrections (captured by INC trades) tend to reflect systematic overcommitment in the day-ahead market, while upward spikes are more episodic and driven by transient system stress. Second, the temporal concentration of profits, with approximately 85\% of total P\&L generated in 2025 during periods of extreme system stress, is less suggestive of a stable mechanical arbitrage and more consistent with compensation for bearing tail risk during stressed system conditions. Together, these observations suggest that the documented profits are consistent with an equilibrium interpretation, rather than evidence of a persistent market inefficiency.}

\revision{A concrete illustration of this interpretation arises during the June 2025 NYISO heat-wave episode, one of the most profitable periods in our out-of-sample backtest. On those days, high demand, reserve tightness, and localized congestion generated unusually large DART dislocations, especially in constrained downstate zones. The model performed well precisely because it assigned elevated spike probabilities to the relevant zone-direction combinations under stressed system conditions. We therefore interpret these profits not as pure arbitrage rents, but as compensation for correctly anticipating short-term pricing risk in a constrained operating environment. This interpretation is further supported by the practical frictions inherent to virtual bidding: participation requires detailed knowledge of market design, access to high-frequency data, and the ability to manage collateral and risk constraints, all of which limit the pool of participants capable of exploiting such signals and help explain the persistence of predictable price dislocations.}

\paragraph{Extensions.} \revision{Several extensions of this study offer promising directions for further research. As noted throughout, the central contribution of this paper lies in the joint treatment of spike forecasting and price-impact-aware position sizing. While we have explored several modeling choices for each component, including alternative classification methods and robustness checks on the impact calibration, our primary objective was not to optimize each component in isolation but to develop a unified and economically consistent framework. Both components nonetheless offer independent directions for further refinement.}

\revision{On the forecasting side, alternative approaches for rare-event prediction, such as resampling techniques (e.g., SMOTE), cost-sensitive learning, or class-weight adjustments, could be explored to better capture extreme price spikes. The economic value of the spike forecasts could also be investigated through alternative instruments beyond virtual INC/DEC trades, such as Financial Transmission Rights (FTRs) or other hedging products that monetize congestion patterns.}

\revision{On the price impact side, the backtesting framework could be enriched by replacing the linear impact proxy with realized day-ahead price perturbations computed directly from historical bid stacks. More flexible structural specifications, such as piecewise-linear approximations, nonlinear bid-stack models, or regime-switching formulations, could capture the dependence of price sensitivity on system conditions, particularly during periods of congestion or scarcity. Incorporating stochastic or time-varying supply curve slopes would further allow the model to reflect changing market conditions across seasons and load regimes.}

\revision{Finally, extending the framework to a multi-player setting, in which virtual traders interact strategically and may condition their bidding behavior on observed price patterns, would provide a richer account of competitive dynamics in two-settlement markets and help assess the long-run sustainability of the documented trading opportunities. Together, these extensions would provide a richer representation of market microstructure and further connect predictive signals with price formation in electricity markets.}

\bibliographystyle{plain}
\bibliography{reference}

\newpage

\appendix


\section{Data Summary Across ISOs}\label{app:data_summary}
\begin{table}[H]
\centering\caption{Summary of sample periods and modeling choices across ISOs.}
\small
\setlength{\tabcolsep}{4pt}
\renewcommand{\arraystretch}{1.1}
\resizebox{\textwidth}{!}{
\begin{tabular}{lccc}
\hline
 & NYISO & ISO--NE & ERCOT \\
\hline
Sample period & 2015--2025 & 2018--2025 & 2018--2025 \\
Train / Val / Test & 2015--19 / 20--21 / 22--25 & 2018--22 / 2023 / 2024--25 & 2018--22 / 2023 / 2024--25 \\
Zones used & 11 (6 highlighted) & 1 (ME) & 1 (WEST) \\
Prediction model & Logistic regression & Logistic regression & Logistic regression \\
Thresholds $(\gamma_{\text{pos}}, \gamma_{\text{neg}})$ & (5, 30) & (2, 8) & (15, 10) \\
Prediction horizon & 19--42h & 13--37h & 14--37h \\
Multi-zone modeling & Yes & No & No \\
\hline
\end{tabular}
}
\label{tab:iso_summary}
\end{table}

\section{Tables and Figures for NYISO}

\begin{figure}[H]
  \centering
  \begin{subfigure}[H]{0.49\textwidth}
    \centering
    \includegraphics[width=\linewidth,height=0.13\textheight,keepaspectratio]{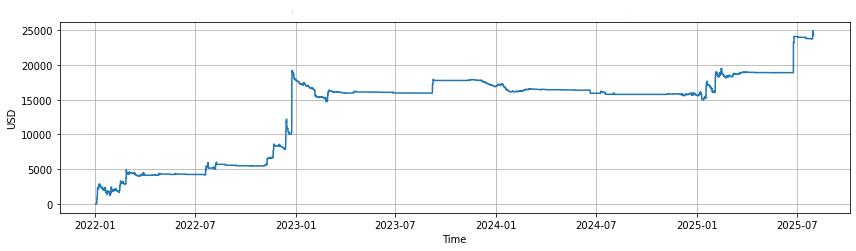}
    \caption{CAPITL — DEC}
  \end{subfigure}\hfill
  \begin{subfigure}[H]{0.49\textwidth}
    \centering
    \includegraphics[width=\linewidth,height=0.13\textheight,keepaspectratio]{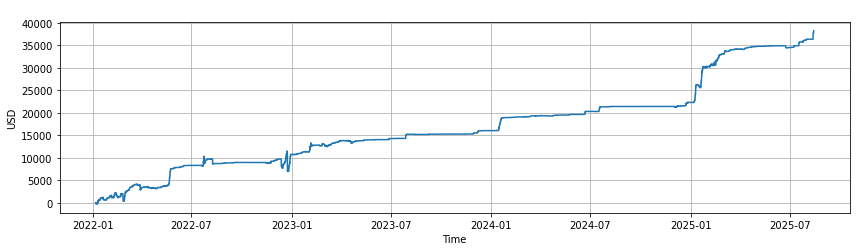}
    \caption{CAPITL — INC}
  \end{subfigure}

  \vspace{0.6em}

  \begin{subfigure}[t]{0.49\textwidth}
    \centering
    \includegraphics[width=\linewidth,height=0.13\textheight,keepaspectratio]{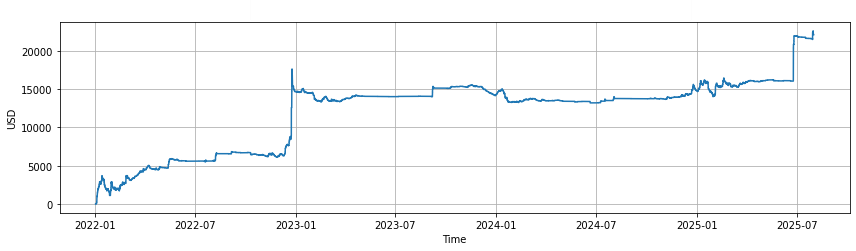}
    \caption{CENTRL — DEC}
  \end{subfigure}\hfill
  \begin{subfigure}[t]{0.49\textwidth}
    \centering
    \includegraphics[width=\linewidth,height=0.13\textheight,keepaspectratio]{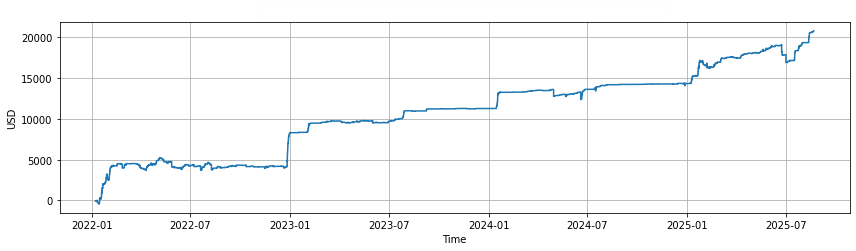}
    \caption{CENTRL — INC}
  \end{subfigure}

  \vspace{0.6em}

  \begin{subfigure}[t]{0.49\textwidth}
    \centering
    \includegraphics[width=\linewidth,height=0.13\textheight,keepaspectratio]{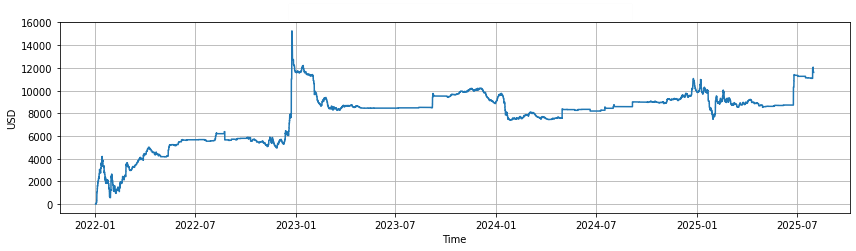}
    \caption{NORTH — DEC}
  \end{subfigure}\hfill
  \begin{subfigure}[t]{0.49\textwidth}
    \centering
    \includegraphics[width=\linewidth,height=0.13\textheight,keepaspectratio]{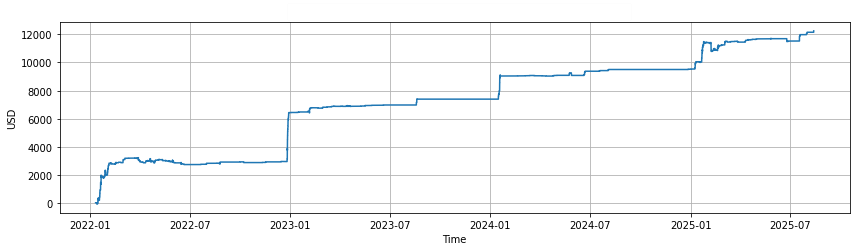}
    \caption{NORTH — INC}
  \end{subfigure}

  \vspace{0.6em}

  \begin{subfigure}[t]{0.49\textwidth}
    \centering
    \includegraphics[width=\linewidth,height=0.13\textheight,keepaspectratio]{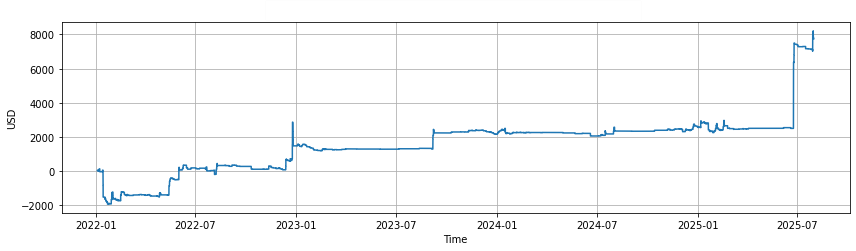}
    \caption{WEST — DEC}
  \end{subfigure}\hfill
  \begin{subfigure}[t]{0.49\textwidth}
    \centering
    \includegraphics[width=\linewidth,height=0.13\textheight,keepaspectratio]{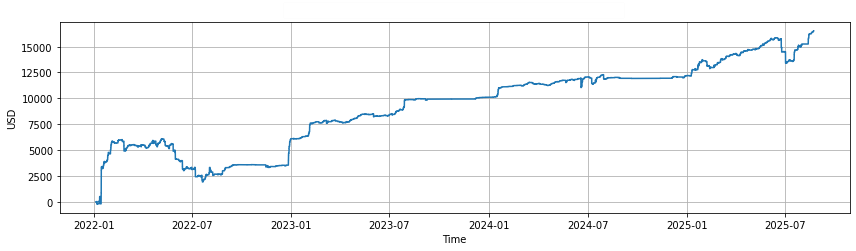}
    \caption{WEST — INC}
  \end{subfigure}

  \caption{NYISO: cumulative P\&L by zone for remaining regions under the INC/DEC benchmark strategy.}
  \label{fig:zone-pnls-appendix}
\end{figure}

\begin{table}[H]
\centering
\small
\caption{NYISO: per-year P\&L by zone (DEC-only benchmark strategy)}
\begin{tabular}{lrrrrrr}
\toprule
Year/Metric & CAPITL & CENTRL & LONGIL & NORTH & NYC & WEST \\
\midrule
2022 & 17,855 & 14,673 & 22,315 & 11,632 & 16,254 & 1,457 \\
2023 &   -886 &   -353 &  3,163 & -2,559 & -2,582 &   679 \\
2024 & -1,383 &    371 &  3,065 &    754 &  3,966 &   454 \\
2025 &  8,731 &  7,409 & 24,445 &  1,776 & 23,035 & 5,162 \\
\midrule
Total & 24,316 & 22,099 & 52,988 & 11,602 & 40,673 & 7,752 \\
Standard Deviation & 6,238 & 7,183 & 10,158 & 6,483 & 6,579 & 3,362 \\
Sharpe ratio & 0.91 & 0.82 & 1.26 & 1.01 & 1.20 & 0.87 \\
\bottomrule
\end{tabular}
\label{tab:zone-pnls-inc}
\end{table}

\begin{table}[H]
\centering
\small
\caption{NYISO: per-year P\&L by zone (INC-only benchmark strategy)}
\label{tab:zone-pnls-dec}
\begin{tabular}{lrrrrrr}
\toprule
Year/Metric & CAPITL & CENTRL & LONGIL & NORTH & NYC & WEST \\
\midrule
2022 & 10,749 &  8,287 &  6,879 &  6,427 &  8,705 &  6,082 \\
2023 &  5,299 &  2,970 &  3,035 &    960 &  3,782 &  4,011 \\
2024 &  6,288 &  3,057 &  5,020 &  2,135 &  2,425 &  2,100 \\
2025 & 15,880 &  6,454 &  1,769 &  2,728 &  1,932 &  4,322 \\
\midrule
Total & 38,215 & 20,768 & 16,703 & 12,250 & 16,844 & 16,516 \\
Standard deviation & 5,439 & 6,143 & 5,301 & 2,850 & 7,507 & 5,276 \\
Sharpe ratio & 1.76 & 0.85 & 0.79 & 1.07 & 0.56 & 0.78 \\
\bottomrule
\end{tabular}
\end{table}

\begin{table}[H]
\centering
\small
\caption{NYISO: yearly mean of DART spreads by zone}
\label{tab:dart-means}
\begin{tabular}{lrrrrrr}
\toprule
Year & CAPITL & CENTRL & LONGIL & NORTH & NYC & WEST \\
\midrule
2015 & 0.33 & 0.27 & 1.09 & 0.25 & 0.54 & -1.82 \\
2016 & 0.67 & 0.20 & 0.67 & 0.25 & -0.02 & -0.13 \\
2017 & 0.47 & -0.11 & 1.40 & 1.27 & 0.90 & -0.31 \\
2018 & -0.17 & -0.49 & 0.80 & -0.54 & -1.02 & -0.01 \\
2019 & 1.06 & 0.48 & -0.56 & -0.11 & 0.71 & -0.29 \\
2020 & -0.08 & -0.10 & -0.08 & -0.60 & -0.60 & -0.03 \\
2021 & 0.40 & 0.33 & -0.84 & -0.79 & 0.03 & 0.08 \\
2022 & -2.51 & -2.10 & -2.37 & -1.56 & -2.80 & 0.58 \\
2023 & 1.76 & 0.29 & -0.29 & 1.02 & 0.47 & 1.27 \\
2024 & 1.38 & 0.54 & 1.20 & 0.06 & -0.09 & 0.95 \\
2025 & 1.62 & -0.68 & -3.18 & -0.14 & -4.05 & 0.50 \\
\bottomrule
\end{tabular}
\end{table}

\section{DART Correlation Across Zones}

\begin{table}[H]\centering
\caption{Correlation of DART {spikes} across NYISO zones (2015--2025)}
\label{tab:dart-corr}
\setlength{\tabcolsep}{2pt}
\scriptsize
\resizebox{\linewidth}{!}{
\begin{tabular}{l
                S[table-format=1.3]
                S[table-format=1.3]
                S[table-format=1.3]
                S[table-format=1.3]
                S[table-format=1.3]
                S[table-format=1.3]
                S[table-format=1.3]
                S[table-format=1.3]
                S[table-format=1.3]
                S[table-format=1.3]
                S[table-format=1.3]}
\toprule
       & {CAPITL} & {CENTRL} & {DUNWOD} & {GENESE} & {HUDVL} & {LONGIL} & {MHKVL} & {MILLWD} & {NORTH} & {NYC} & {WEST} \\
\midrule
CAPITL & 1.000 & 0.710 & 0.778 & 0.688 & 0.867 & 0.603 & 0.718 & 0.834 & 0.465 & 0.754 & 0.562 \\
CENTRL & 0.710 & 1.000 & 0.776 & 0.987 & 0.864 & 0.618 & 0.990 & 0.827 & 0.737 & 0.755 & 0.810 \\
DUNWOD & 0.778 & 0.776 & 1.000 & 0.759 & 0.904 & 0.747 & 0.780 & 0.966 & 0.533 & 0.965 & 0.609 \\
GENESE & 0.688 & 0.987 & 0.759 & 1.000 & 0.844 & 0.606 & 0.978 & 0.808 & 0.728 & 0.738 & 0.797 \\
HUDVL  & 0.867 & 0.864 & 0.904 & 0.844 & 1.000 & 0.713 & 0.864 & 0.960 & 0.594 & 0.879 & 0.679 \\
LONGIL & 0.603 & 0.618 & 0.747 & 0.606 & 0.713 & 1.000 & 0.618 & 0.744 & 0.415 & 0.726 & 0.484 \\
MHKVL  & 0.718 & 0.990 & 0.780 & 0.978 & 0.864 & 0.618 & 1.000 & 0.829 & 0.784 & 0.758 & 0.786 \\
MILLWD & 0.834 & 0.827 & 0.966 & 0.808 & 0.960 & 0.744 & 0.829 & 1.000 & 0.570 & 0.935 & 0.649 \\
NORTH  & 0.465 & 0.737 & 0.533 & 0.728 & 0.594 & 0.415 & 0.784 & 0.570 & 1.000 & 0.519 & 0.557 \\
NYC    & 0.754 & 0.755 & 0.965 & 0.738 & 0.879 & 0.726 & 0.758 & 0.935 & 0.519 & 1.000 & 0.593 \\
WEST   & 0.562 & 0.810 & 0.609 & 0.797 & 0.679 & 0.484 & 0.786 & 0.649 & 0.557 & 0.593 & 1.000 \\
\bottomrule
\end{tabular}
}
\end{table}

\begin{table}[H]
\centering
\caption{Correlation of DART {spreads} across ISO--NE zones (2018--2025).}
\label{tab:isone_dart_corr}
\begin{tabular}{lcccccccc}
\toprule
\textbf{Region} & CT & ME & NEMASS & NH & RI & SEMASS & VT & WCMASS \\
\midrule
CT          & 1.000 & 0.988 & 0.995 & 0.996 & 0.995 & 0.995 & 0.999 & 0.998 \\
ME          & 0.988 & 1.000 & 0.993 & 0.994 & 0.993 & 0.993 & 0.990 & 0.993 \\
NEMASS      & 0.995 & 0.993 & 1.000 & 0.999 & 0.999 & 0.999 & 0.995 & 0.998 \\
NH          & 0.996 & 0.994 & 0.999 & 1.000 & 0.999 & 0.999 & 0.997 & 0.999 \\
RI          & 0.995 & 0.993 & 0.999 & 0.999 & 1.000 & 1.000 & 0.996 & 0.999 \\
SEMASS      & 0.995 & 0.993 & 0.999 & 0.999 & 1.000 & 1.000 & 0.996 & 0.999 \\
VT          & 0.999 & 0.990 & 0.995 & 0.997 & 0.996 & 0.996 & 1.000 & 0.999 \\
WCMASS      & 0.998 & 0.993 & 0.998 & 0.999 & 0.999 & 0.999 & 0.999 & 1.000 \\
\bottomrule
\end{tabular}
\end{table} 
\begin{table}[H]
\centering
\caption{Correlation matrix of DART across ERCOT zones (2018--2025).}
\begin{tabular}{lcccc}
\hline
\textbf{Zone} & \textbf{NORTH} & \textbf{SOUTH} & \textbf{WEST} & \textbf{HOUSTON} \\
\hline
NORTH   & 1.000 & 0.991 & 0.998 & 0.980 \\
SOUTH   & 0.991 & 1.000 & 0.991 & 0.981 \\
WEST    & 0.998 & 0.991 & 1.000 & 0.978 \\
HOUSTON & 0.980 & 0.981 & 0.978 & 1.000 \\
\hline
\end{tabular}
\label{tab:ercot-dart-corr}
\end{table}

\section{NYISO, ISO--NE \& ERCOT Quantiles}
\begin{table}[H]
\centering
\caption{Empirical DART quantiles (USD/MWh) on the training sets for NYISO LONGIL, ISO--NE ME, and ERCOT WEST.}
\label{tab:dart-quantiles-longil-maine-west}
\small
\setlength{\tabcolsep}{10pt}
\begin{tabular}{lrrr}
\toprule
\textbf{Quantile} & \textbf{LONGIL (NYISO)} & \textbf{ME (ISO--NE)} & \textbf{WEST (ERCOT)} \\
\midrule
$Q_{0.00}$ & -2434.97 & -109.17 & -5891.08 \\
$Q_{0.01}$ &  -121.57 &  -29.60 &   -77.20 \\
$Q_{0.05}$ &   -34.95 &  -13.39 &   -13.74 \\
$Q_{0.10}$ &   -16.33 &   -7.10 &    -7.65 \\
$Q_{0.25}$ &    -3.09 &   -1.51 &    -2.02 \\
$Q_{0.50}$ &     3.98 &    1.33 &     0.97 \\
$Q_{0.75}$ &    11.21 &    4.62 &     4.96 \\
$Q_{0.90}$ &    20.27 &    9.31 &    11.39 \\
$Q_{0.95}$ &    29.18 &   14.03 &    17.39 \\
$Q_{0.99}$ &    58.25 &   27.75 &    40.39 \\
$Q_{1.00}$ &  1506.76 &   69.03 &  7675.51 \\
\bottomrule
\end{tabular}
\end{table}

\section{Figures and Tables for the Optimal Trading Strategy}
This subsection collects supplementary calibration results, diagnostic plots, and
robustness checks underlying the trading strategy in Section~\ref{sec:scaling-impact}.
It includes (i) price–impact estimates by season and load band,
(ii) validation and test–set execution diagnostics,
and (iii) additional performance breakdowns and distributional comparisons.

\begin{table}[H]
\centering
\begin{threeparttable}
\caption{(Loss -- Congestion) impact on LMP for a +1000 MW zonal load change (\$/MWh), 2015--2021.}
\label{tab:impact-loss-minus-cong-new}
\small
\begin{tabular}{l*{6}{S[table-figures-decimal=2]}}
\toprule
& \multicolumn{2}{c}{Shoulder} & \multicolumn{2}{c}{Summer} & \multicolumn{2}{c}{Winter} \\
\cmidrule(lr){2-3}\cmidrule(lr){4-5}\cmidrule(lr){6-7}
Zone & {Off-Peak} & {Peak} & {Off-Peak} & {Peak} & {Off-Peak} & {Peak} \\
\midrule
CAPITL &  8.56 &  2.36 &  -7.86 &  0.86 &  63.41 &  79.54 \\
CENTRL &  2.61 &  2.87 &  -0.37 &  2.39 &   4.16 &   4.56 \\
DUNWOD &  3.61 &  9.36 &  -5.20 & 11.39 & 111.28 & 104.13 \\
GENESE &  1.76 &  1.70 &  -0.41 &  1.21 &   1.49 &   0.04 \\
HUDVL  &  9.51 &  5.91 &  -3.51 &  5.04 &  68.75 &  65.51 \\
LONGIL &  4.95 &  7.82 &   5.06 & 17.73 &  43.63 &  43.30 \\
MHKVL  &  5.30 &  4.86 &  -0.37 &  4.21 &  13.18 &  18.27 \\
MILLWD & 52.70 & 46.17 & -11.37 & 25.69 & 189.06 & 192.38 \\
NORTH  &  0.26 &  3.80 & -19.55 & -44.72 &  -2.82 &  -0.91 \\
NYC    & -0.25 &  1.29 &  -0.05 &  2.18 &  13.21 &  12.29 \\
WEST   &  1.88 & 14.36 &  -0.54 & 25.73 &   1.54 &   1.11 \\
\bottomrule
\end{tabular}
\end{threeparttable}
\end{table}

\begin{table}[H]
\centering
\begin{threeparttable}
\caption{Average forecast load (MW) by zone, season, and Peak/Off-Peak, 2015--2021.}
\label{tab:avg-load-2015-2021}
\small
\begin{tabular}{l*{6}{S[round-precision=1,table-figures-decimal=1]}}
\toprule
& \multicolumn{2}{c}{Shoulder} & \multicolumn{2}{c}{Summer} & \multicolumn{2}{c}{Winter} \\
\cmidrule(lr){2-3}\cmidrule(lr){4-5}\cmidrule(lr){6-7}
Zone & {Off-Peak} & {Peak} & {Off-Peak} & {Peak} & {Off-Peak} & {Peak} \\
\midrule
CAPITL & 1133.2 & 1336.2 & 1330.3 & 1645.6 & 1294.1 & 1486.9 \\
CENTRL & 1503.0 & 1785.7 & 1629.5 & 2041.0 & 1728.3 & 2003.5 \\
DUNWOD &  542.8 &  676.9 &  706.1 &  906.1 &  597.4 &  720.3 \\
GENESE &  918.8 & 1124.1 & 1057.9 & 1367.4 & 1023.3 & 1220.3 \\
HUDVL  &  879.3 & 1053.3 & 1073.4 & 1376.3 & 1012.0 & 1172.0 \\
LONGIL & 1887.4 & 2327.4 & 2636.2 & 3435.6 & 2061.7 & 2462.3 \\
MHKVL  &  632.3 &  767.0 &  674.4 &  866.9 &  786.5 &  925.6 \\
MILLWD &  250.4 &  309.6 &  309.2 &  404.6 &  305.3 &  362.5 \\
NORTH  &  499.4 &  532.5 &  484.8 &  531.3 &  586.6 &  619.9 \\
NYC    & 4819.9 & 6102.0 & 6307.2 & 7926.5 & 5074.9 & 6249.1 \\
WEST   & 1479.6 & 1734.2 & 1637.3 & 2004.4 & 1624.4 & 1872.3 \\
\bottomrule
\end{tabular}
\end{threeparttable}
\end{table}

\begin{figure}[H]
    \centering
    \includegraphics[width=0.6\textwidth]{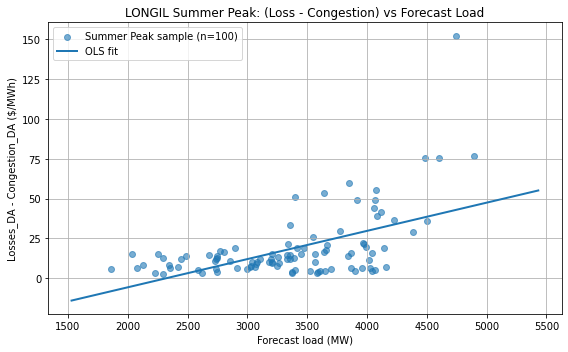}
    \caption{Scatter of (Forecasted Load,{\text{Loss}+\text{Congestion}}) for 100 random Summer--Peak hours in LONGIL.}
    \label{fig:random-100}
\end{figure}

\begin{table}[H]
\centering
\small
\caption{Average correlations across NYISO zones between forecast load and DA congestion/loss components, by season and Peak/Off-Peak bucket (2015--2021).}
\label{tab:avg_corr_load_cong_loss}
\begin{tabular}{lcc}
\toprule
\textbf{Season -- Bucket} 
& \textbf{Corr(Forecast Load, Losses)} 
& \textbf{Corr(Forecast Load, Congestion)} \\
\midrule
Shoulder -- Off-Peak & 0.317 & $0.050$ \\
Shoulder -- Peak    & 0.373 & $0.073$ \\
Summer -- Off-Peak   & 0.352 &  0.112 \\
Summer -- Peak      & 0.397 & $0.177$ \\
Winter -- Off-Peak   & 0.314 & $0.239$ \\
Winter -- Peak      & 0.274 & $0.238$ \\
\bottomrule
\end{tabular}
\end{table}

\begin{table}[H]
\centering
\small
\setlength{\tabcolsep}{6pt}
\caption{Execution-view P\&L by zone, aggregated over 2022--2025 (USD).}
\label{tab:zone_exec_agg}
\begin{tabular}{lrrr}
\toprule
Zone & DEC (Exec) & INC (Exec) & Total \\
\midrule
CAPITL &  -6,564 & -349,354 & -355,918 \\
CENTRL &  -3,052 & -206,313 & -209,365 \\
DUNWOD & -38,177 & -131,202 & -169,379 \\
GENESE &  49,801 &  69,174 & 119,975 \\
HUDVL  &  -3,023 & -411,510 & -414,533 \\
LONGIL & 6,004,479 &       0 & 6,004,479 \\
MHKVL  &  -1,178 & -101,653 & -102,831 \\
MILLWD & 128,091 &   -8,946 & 119,145 \\
NORTH  & -14,361 &   3,556 & -10,805 \\
NYC    & 778,698 & -56,080 & 722,618 \\
WEST   &   685 &  26,153 & 26,838 \\
\bottomrule
\end{tabular}
\end{table}

\begin{table}[H]
\centering
\small
\caption{Yearly total P\&L by view and side (USD).}
\label{tab:yearly_totals_small}
\begin{tabular}{rrrrrrr}
\toprule
Year & DEC (Exec) & INC (Exec)   & DEC (Pred) & INC (Pred) & Total (Pred) \\
\midrule
2022 & 915{,}311 & -201{,}181  & 411{,}226 & 302{,}904 & 714{,}130 \\
2023 & 336{,}339 &  -93{,}707   & 198{,}424 &  44{,}207 & 242{,}631 \\
2024 & -185{,}819 & 296{,}156  & -68{,}264 & 178{,}601 & 110{,}337 \\
2025 & 5{,}829{,}911 & -1{,}136{,}856 & 4{,}560{,}358 & 132{,}698 & 4{,}693{,}055 \\
\bottomrule
\end{tabular}
\end{table}

\begin{table}[H]
\centering
\caption{Realized day-ahead price impact on 24 June 2025 for the executed net DEC portfolio and the corresponding DART in Long Island (NYISO, upward shift of residual demand by $q_t$ MWh).}
\label{tab:impact-24jun-2025}
\small
\begin{tabular}{
  l
  S[table-format=3.1]  
  S[table-format=3.2]  
  S[table-format=4.2]  
  S[table-format=2.2]  
  S[table-format=2.2]  
  S[table-format=7.2]  
}
\toprule
\textbf{Hour}
& {$q_t$}
& {$P^{\mathrm{DA}}_t$}
& {$P^{\mathrm{DA}}_t(q_t)$}
& {$\Delta P_t$ (\$/MWh)}
& {$1000\,\Delta P_t / q_t$}
& {DART} \\
\midrule
15{:}00 &  72.6 & 209.41 & 210.92 & 1.51 & 20.80 & -1159.57 \\
16{:}00 &  85.9 & 222.59 & 224.79 & 2.20 & 25.61 &  -934.65 \\
17{:}00 & 159.9 & 250.58 & 255.90 & 5.32 & 33.27 & -3725.93 \\
18{:}00 & 163.1 & 224.79 & 226.59 & 1.80 & 11.04 & -4372.52 \\
19{:}00 & 163.8 & 201.72 & 203.14 & 1.42 &  8.67 & -1798.57 \\
20{:}00 & 161.3 & 170.00 & 176.57 & 6.57 & 40.73 &  -534.26 \\
21{:}00 & 134.2 & 153.00 & 155.00 & 2.00 & 14.90 &  -145.64 \\
\bottomrule
\end{tabular}
\end{table}

\begin{figure}[ht]
    \centering
    \includegraphics[width=0.8\linewidth]{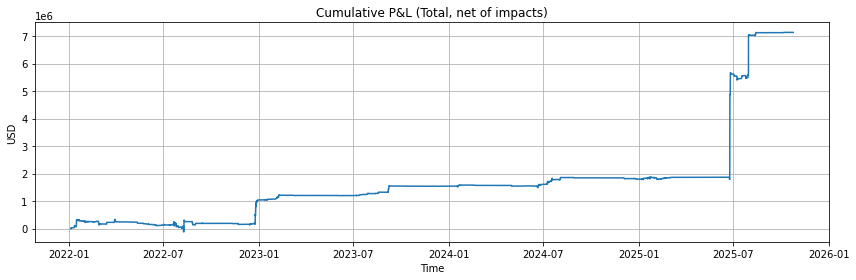}
    \caption{Cumulative P\&L under regression-based impact coefficients $(k_z)$.}
    \label{fig:pnl_new_k}
\end{figure}
\begin{table}[ht]
\centering
\small
\caption{TEST precision/recall/F1 by zone for DEC predicted.}
\label{tab:test_zone_metrics_inc}
\setlength{\tabcolsep}{6pt}
\begin{tabular}{lcccrrrrrr}
\hline
zone & precision & recall & f1 & TP & FP & FN & TN & support\_pos & N \\
\hline
LONGIL & 0.270 & 0.066 & 0.107 & 154 & 417 & 2167 & 29250 & 2321 & 31988 \\
WEST   & 0.216 & 0.063 & 0.097 &  71 & 257 & 1064 & 30572 & 1135 & 31964 \\
NYC    & 0.291 & 0.047 & 0.080 &  75 & 183 & 1535 & 30171 & 1610 & 31964 \\
MILLWD & 0.323 & 0.042 & 0.075 &  63 & 132 & 1430 & 31851 & 1493 & 33476 \\
CAPITL & 0.309 & 0.041 & 0.072 &  73 & 163 & 1714 & 30014 & 1787 & 31964 \\
HUDVL  & 0.302 & 0.041 & 0.072 &  57 & 132 & 1347 & 31940 & 1404 & 33476 \\
DUNWOD & 0.309 & 0.040 & 0.071 &  60 & 134 & 1433 & 31849 & 1493 & 33476 \\
CENTRL & 0.310 & 0.035 & 0.063 &  45 & 100 & 1230 & 30589 & 1275 & 31964 \\
GENESE & 0.314 & 0.034 & 0.062 &  43 &  94 & 1206 & 32133 & 1249 & 33476 \\
MHKVL  & 0.295 & 0.033 & 0.060 &  44 & 105 & 1276 & 32051 & 1320 & 33476 \\
NORTH  & 0.343 & 0.030 & 0.056 &  49 &  94 & 1564 & 30281 & 1613 & 31988 \\
\hline
\end{tabular}
\end{table}
\begin{table}[ht]
\centering
\small
\caption{TEST precision/recall/F1 by zone for INC predicted.}
\label{tab:test_zone_metrics_dec}
\setlength{\tabcolsep}{6pt}
\begin{tabular}{lcccrrrrrr}
\hline
zone & precision & recall & f1 & TP & FP & FN & TN & support\_pos & N \\
\hline
LONGIL & 0.645 & 0.146 & 0.238 & 2035 & 1118 & 11902 & 16933 & 13937 & 31988 \\
WEST   & 0.747 & 0.059 & 0.110 &  635 &  215 & 10086 & 21028 & 10721 & 31964 \\
CAPITL & 0.795 & 0.052 & 0.098 &  635 &  164 & 11584 & 19581 & 12219 & 31964 \\
NYC    & 0.722 & 0.050 & 0.093 &  560 &  216 & 10723 & 20465 & 11283 & 31964 \\
DUNWOD & 0.782 & 0.040 & 0.077 &  480 &  134 & 11386 & 21476 & 11866 & 33476 \\
MILLWD & 0.776 & 0.038 & 0.073 &  453 &  131 & 11383 & 21509 & 11836 & 33476 \\
HUDVL  & 0.801 & 0.037 & 0.071 &  442 &  110 & 11466 & 21458 & 11908 & 33476 \\
CENTRL & 0.786 & 0.023 & 0.046 &  235 &   64 &  9795 & 21870 & 10030 & 31964 \\
MHKVL  & 0.796 & 0.020 & 0.040 &  223 &   57 & 10675 & 22521 & 10898 & 33476 \\
GENESE & 0.792 & 0.020 & 0.039 &  209 &   55 & 10338 & 22874 & 10547 & 33476 \\
NORTH  & 0.776 & 0.013 & 0.025 &  125 &   36 &  9725 & 22102 &  9850 & 31988 \\
\hline
\end{tabular}
\end{table}

\begin{figure}[H]
  \centering
  \begin{subfigure}{0.48\linewidth}
    \centering
    \includegraphics[width=\linewidth]{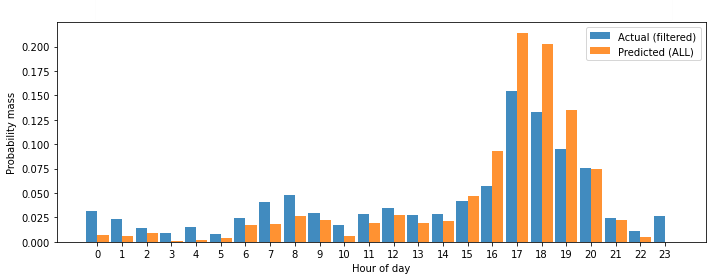}
    \caption{DEC --- Actual top 20\% vs.\ predicted}
    \label{fig:hour_inc_top20}
  \end{subfigure}\hfill
  \begin{subfigure}{0.48\linewidth}
    \centering
    \includegraphics[width=\linewidth]{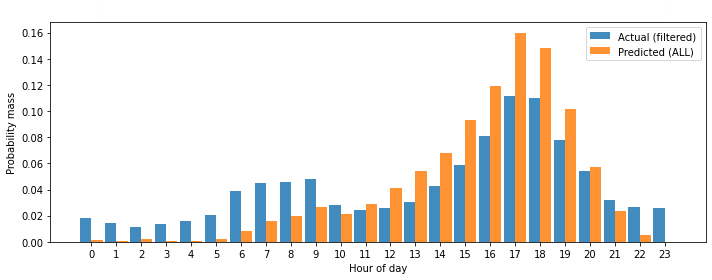}
    \caption{INC --- Actual top 5\% vs.\ predicted}
    \label{fig:hour_dec_top05}
  \end{subfigure}

  \vspace{0.5em}

  \begin{subfigure}{0.48\linewidth}
    \centering
    \includegraphics[width=\linewidth]{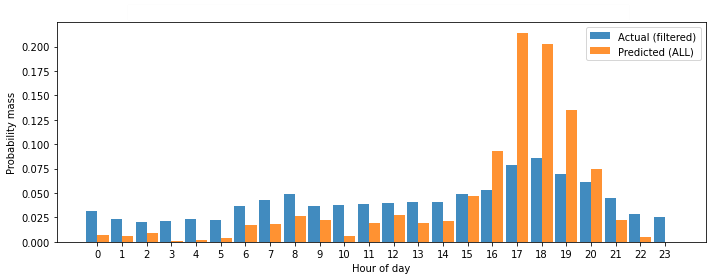}
    \caption{DEC --- All hours vs.\ predicted}
    \label{fig:hour_inc_all}
  \end{subfigure}\hfill
  \begin{subfigure}{0.48\linewidth}
    \centering
    \includegraphics[width=\linewidth]{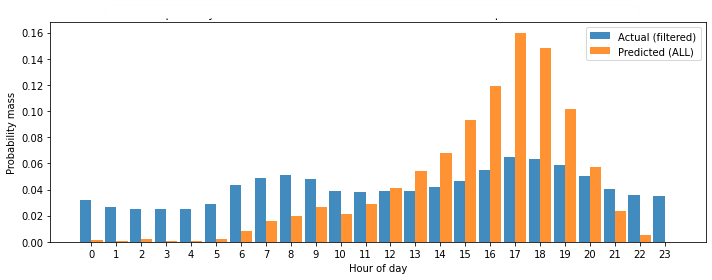}
    \caption{INC --- All hours vs.\ predicted}
    \label{fig:hour_dec_all}
  \end{subfigure}

  \caption{Predicted vs.\ realized spike PDFs across \textbf{hour of day}
  (test period 2022--2025).}
  \label{fig:hour_histograms}
\end{figure}

\begin{figure}[H]
  \centering
  \begin{subfigure}{0.48\linewidth}
    \centering
    \includegraphics[width=\linewidth]{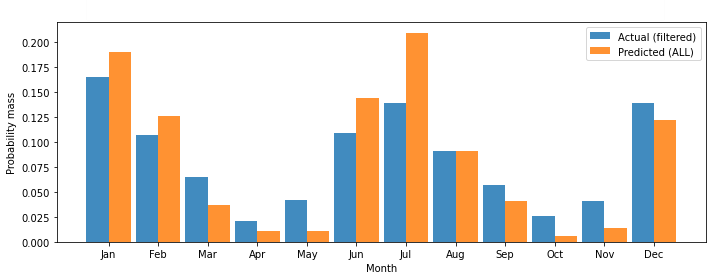}
    \caption{DEC --- Actual top 20\% vs.\ predicted}
    \label{fig:month_inc_top20}
  \end{subfigure}\hfill
  \begin{subfigure}{0.48\linewidth}
    \centering
    \includegraphics[width=\linewidth]{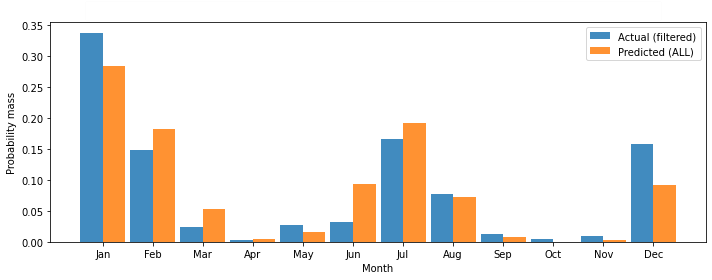}
    \caption{INC --- Actual top 5\% vs.\ predicted}
    \label{fig:month_dec_top05}
  \end{subfigure}

  \vspace{0.5em}

  \begin{subfigure}{0.48\linewidth}
    \centering
    \includegraphics[width=\linewidth]{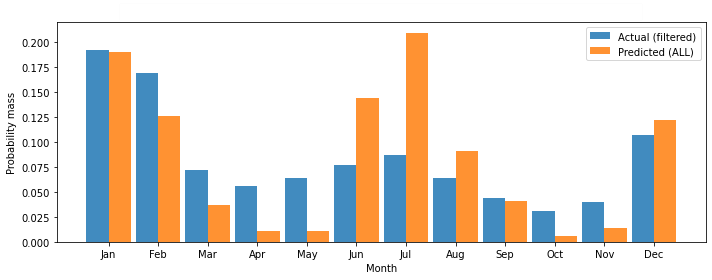}
    \caption{DEC --- All months vs.\ predicted}
    \label{fig:month_inc_all}
  \end{subfigure}\hfill
  \begin{subfigure}{0.48\linewidth}
    \centering
    \includegraphics[width=\linewidth]{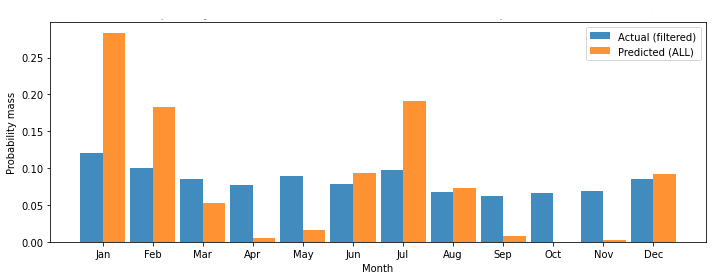}
    \caption{INC --- All months vs.\ predicted}
    \label{fig:month_dec_all}
  \end{subfigure}

  \caption{Predicted vs.\ realized spike PDFs across {month of year}
  (test period 2022--2025).}
  \label{fig:month_histograms}
\end{figure}
\begin{figure}[H]
  \centering
  \begin{subfigure}{0.32\textwidth}
    \centering
    \includegraphics[height=0.15\textheight,width=\textwidth]{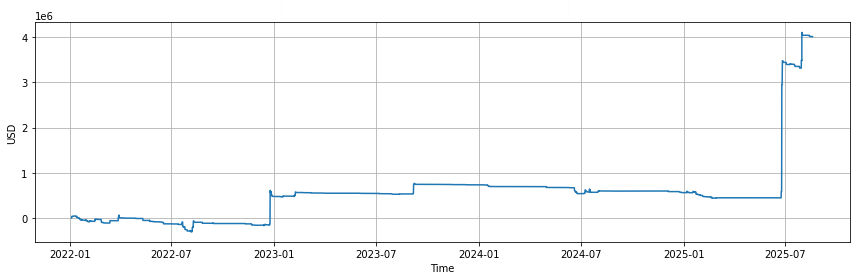}
    \caption{DEC}
    \label{fig:inc-new}
  \end{subfigure}
  \hfill
  \begin{subfigure}{0.32\textwidth}
    \centering
    \includegraphics[height=0.15\textheight,width=\textwidth]{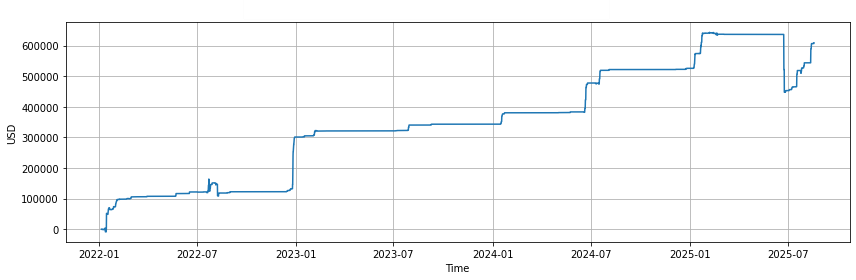}
    \caption{INC}
    \label{fig:dec-new}
  \end{subfigure}
  \hfill
  \begin{subfigure}{0.32\textwidth}
    \centering
    \includegraphics[height=0.15\textheight,width=\textwidth]{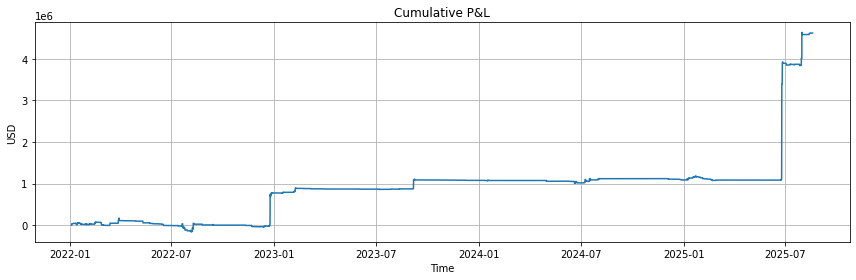}
    \caption{Total}
    \label{fig:pnl-new}
  \end{subfigure}
  \caption{Cumulative P\&L for the restricted (statistically significant) strategy,
  test period 2022--2025.}
  \label{fig:pnl-panels}
\end{figure}

\begin{table}[H]
\centering
\caption{Prediction quality on the 2022--2025 test set for the restricted strategy.}
\label{tab:pred-quality}
\begin{tabular}{lccc}
\toprule
Side & Trades & Spikes & Correct sign \\
\midrule
DEC & 571  & 154 (27.0\%)  & 232 (40.6\%) \\
INC & 2720 & 2065 (75.9\%) & 2173 (79.9\%) \\
\bottomrule
\end{tabular}
\end{table}

\begin{table}[ht]
\centering
\caption{VALID 2020--2021, DEC side: mean P\&L (USD/MWh) and number of trades by zone, season and band.}
\label{tab:inc_valid_zone_season_band}
\begin{tabular}{lrrrrrr}
\toprule
Zone & W P & W O & Su P & Su O & Sh P & Sh O \\
\midrule
CAPITL & 16.24 (11) & --        &  -5.75 (57)  & --        & 28.11 (7)  & -- \\
CENTRL & -3.82 (5)  & --        &  -2.04 (26)  & --        & 30.73 (5)  & -- \\
DUNWOD &  6.43 (7)  & --        &   4.95 (36)  & --        & 25.85 (6)  & -- \\
GENESE & -4.32 (5)  & --        &  17.47 (31)  & --        & -7.44 (4)  & -- \\
HUDVL  & 15.17 (6)  & --        &  -2.50 (36)  & --        & 30.81 (5)  & -- \\
LONGIL & 19.20 (47) & --        &  44.50 (246) & --        &  8.60 (23) & -- \\
MHKVL  & -3.73 (5)  & --        &  -1.32 (24)  & --        & 34.43 (4)  & -- \\
MILLWD &  7.61 (7)  & --        &  19.64 (43)  & --        & 25.95 (6)  & -- \\
NORTH  & -9.10 (6)  & --        &  -2.34 (31)  & --        & 30.43 (4)  & -- \\
NYC    &  4.31 (8)  & --        &  12.30 (107) & --        & 22.37 (7)  & -- \\
WEST   &  3.08 (19) & --        &   5.96 (116) & --        &  2.46 (17) & -- \\
\bottomrule
\end{tabular}
\end{table}

\begin{table}[ht]
\centering
\caption{VALID 2020--2021, INC side: mean P\&L (USD/MWh) and number of trades by zone, season and band.}
\label{tab:dec_valid_zone_season_band}
\begin{tabular}{lrrrrrr}
\toprule
Zone & W P & W O & Su P & Su O & Sh P & Sh O \\
\midrule
CAPITL & 27.94 (50) & --          &  6.24 (149) & --         & -8.64 (7)  & -- \\
CENTRL & 41.40 (7)  & 21.82 (1)   & 14.06 (51)  & --         & 16.52 (2)  & -- \\
DUNWOD & 29.13 (48) & --          & -6.55 (134) & --         &  7.45 (7)  & -- \\
GENESE & 36.22 (5)  & 22.67 (1)   & 11.26 (43)  & --         &  6.71 (2)  & -- \\
HUDVL  & 31.37 (29) & --          &  4.93 (125) & --         & 10.36 (5)  & -- \\
LONGIL &  1.89 (814)& 49.54 (5)   & -7.50 (655) & --         &  5.61 (231)& -- \\
MHKVL  & 43.09 (7)  & 23.22 (1)   & 12.03 (43)  & --         & 14.43 (1)  & -- \\
MILLWD & 28.71 (43) & --          & -9.72 (118) & --         &  8.19 (6)  & -- \\
NORTH  & 28.44 (2)  & 27.61 (1)   &  3.36 (18)  & --         &  4.24 (4)  & -- \\
NYC    & 27.93 (66) & --          & -3.08 (252) & --         &  7.76 (8)  & -- \\
WEST   & 15.06 (81) & 21.58 (1)   &  7.63 (208) & --         &  9.85 (17) & -- \\
\bottomrule
\end{tabular}
\end{table}

\begin{figure}[ht!]
\centering
\includegraphics[width=0.85\textwidth]{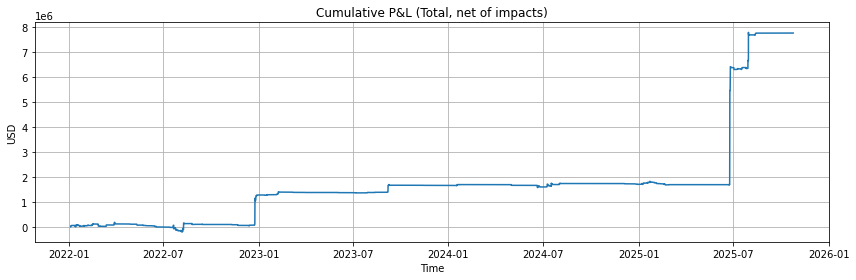}\\[6pt]

\begin{minipage}{0.49\textwidth}
  \centering
  \includegraphics[width=\textwidth]{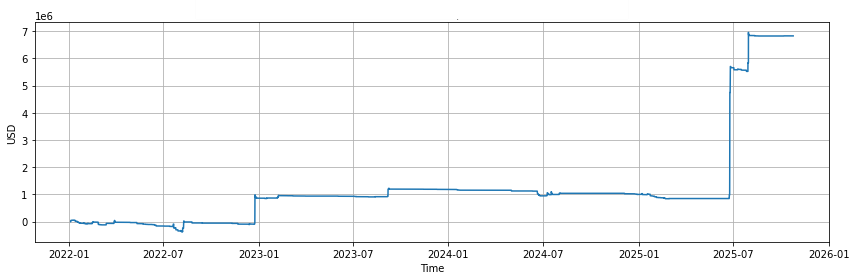}
  \vspace{4pt}
  \small DEC (clipped)
\end{minipage}
\hfill
\begin{minipage}{0.49\textwidth}
  \centering
  \includegraphics[width=\textwidth]{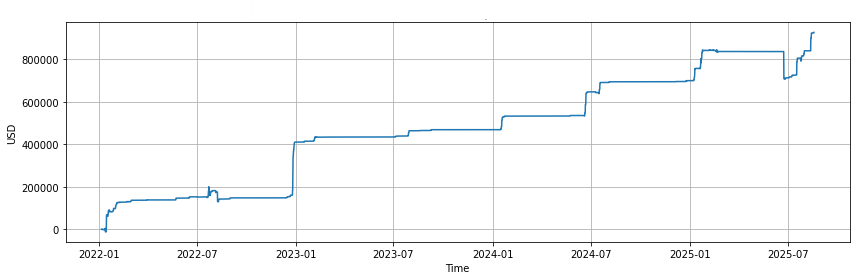}
  \vspace{4pt}
  \small INC (clipped)
\end{minipage}
\caption{Cumulative P\&L with side-frozen (clipped) strategy.  
Top: total portfolio; bottom: DEC and INC contributions.}
\label{fig:clip_results}
\end{figure}
\begin{table}[H]
\centering
\small
\setlength{\tabcolsep}{8pt}
\caption{Per-zone attribution of the clipped strategy on the 2022--2025 test set (execution view).}
\label{tab:perzone_exec_plain}
\begin{tabular}{lrrr}
\hline
\textbf{Zone} & \textbf{Active hours} & \textbf{Avg.\ $|q|$ (MW)} & \textbf{P\&L (USD)} \\
\hline
LONGIL & 570 & 118.91 & 5{,}692{,}023 \\
NYC    & 464 &  21.92 &   910{,}936 \\
CAPITL & 777 &  10.10 &   303{,}353 \\
WEST   & 807 &   9.89 &   242{,}054 \\
CENTRL & 235 &  23.66 &   189{,}883 \\
MILLWD & 184 &   6.23 &   147{,}420 \\
GENESE & 254 &  10.27 &   115{,}452 \\
MHKVL  & 216 &  10.60 &    90{,}960 \\
HUDVL  & 445 &   5.50 &    57{,}758 \\
NORTH  &  28 &   5.93 &     3{,}531 \\
DUNWOD &  57 &   4.06 &    -1{,}143 \\
\hline
\textbf{Total} & 4{,}037 & 26.81 (w.\ avg.) & \textbf{7{,}752{,}227} \\
\hline
\end{tabular}
\end{table}

\end{document}